\documentclass[
twocolumn,
superscriptaddress,
amsmath,amssymb,
aps,
prr,
floatfix,
]{revtex4-2}

\usepackage{mathtools}
\usepackage{amsmath}
\allowdisplaybreaks
\usepackage{ bbold }
\usepackage{amssymb}
\usepackage[normalem]{ulem}
\usepackage{amsthm}
\usepackage{xfrac}
\usepackage{dsfont}
\usepackage{siunitx}
\usepackage{physics}
\usepackage{graphicx}
\usepackage[breaklinks=true,colorlinks,citecolor=blue,linkcolor=blue,urlcolor=blue]{hyperref}
\usepackage[capitalise]{cleveref}
\usepackage{dcolumn}
\usepackage{bm}
\usepackage{layouts}
\usepackage{comment}
\usepackage{soul}
\usepackage[caption=false]{subfig}
\usepackage[dvipsnames]{xcolor}

\newcommand{\phantomsubcaptionlabel}[1]{\refstepcounter{subfigure}\label{#1}}

\newcommand{\fil}[1]{\textcolor{blue}{#1}}

\newcommand{\rmd}{{\rm d}}
\newcommand{\rmi}{{\rm i}}

\begin{document}

\title{Dissipative phase transitions and chaos in two-photon driven quantum optomechanics}

\author{Giovanni Bragadin}
\altaffiliation{These authors contributed equally to this work. Correspondence should be addressed to \fil{filippo.ferrari@epfl.ch}}
\affiliation{Institute of Physics, Ecole Polytechnique Fédérale de Lausanne (EPFL), CH-1015 Lausanne, Switzerland}
\affiliation{Center for Quantum Science and Engineering, Ecole Polytechnique Fédérale de Lausanne (EPFL), CH-1015 Lausanne, Switzerland}

\author{Filippo Ferrari}
\altaffiliation{These authors contributed equally to this work. Correspondence should be addressed to \fil{filippo.ferrari@epfl.ch}}
\affiliation{Institute of Physics, Ecole Polytechnique Fédérale de Lausanne (EPFL), CH-1015 Lausanne, Switzerland}
\affiliation{Center for Quantum Science and Engineering, Ecole Polytechnique Fédérale de Lausanne (EPFL), CH-1015 Lausanne, Switzerland}

\author{Lorenzo Fioroni}
\altaffiliation{These authors contributed equally to this work. Correspondence should be addressed to \fil{filippo.ferrari@epfl.ch}}
\affiliation{Institute of Physics, Ecole Polytechnique Fédérale de Lausanne (EPFL), CH-1015 Lausanne, Switzerland}
\affiliation{Center for Quantum Science and Engineering, Ecole Polytechnique Fédérale de Lausanne (EPFL), CH-1015 Lausanne, Switzerland}

\author{Matteo Seclì}
\affiliation{Institute of Physics, Ecole Polytechnique Fédérale de Lausanne (EPFL), CH-1015 Lausanne, Switzerland}
\affiliation{Center for Quantum Science and Engineering, Ecole Polytechnique Fédérale de Lausanne (EPFL), CH-1015 Lausanne, Switzerland}

\author{Vincenzo Macrì}
\affiliation{Dipartimento di Fisica ``A. Volta”, Università di Pavia, Via Bassi 6, 27100 Pavia, Italy}

\author{Vincenzo Savona}
\affiliation{Institute of Physics, Ecole Polytechnique Fédérale de Lausanne (EPFL), CH-1015 Lausanne, Switzerland}
\affiliation{Center for Quantum Science and Engineering, Ecole Polytechnique Fédérale de Lausanne (EPFL), CH-1015 Lausanne, Switzerland}

\author{Alberto Mercurio}
\affiliation{Institute of Physics, Ecole Polytechnique Fédérale de Lausanne (EPFL), CH-1015 Lausanne, Switzerland}
\affiliation{Center for Quantum Science and Engineering, Ecole Polytechnique Fédérale de Lausanne (EPFL), CH-1015 Lausanne, Switzerland}

\date{\today}

\begin{abstract}
We investigate nonequilibrium criticality and chaos in a two-photon-driven optomechanical system. The parametric drive preserves a discrete $\mathbb{Z}_2$ symmetry of the optical field, while radiation-pressure coupling transfers the resulting nonlinear dynamics to a mechanical oscillator. Combining semiclassical stability analysis, exact Liouvillian spectra, and stochastic quantum trajectories, we show that this driven-dissipative optomechanical model supports both first- and second-order dissipative phase transitions. At negative detuning a second-order transition yields spontaneous breaking of the cavity-parity symmetry in the thermodynamic limit. At positive detuning the same symmetry breaking coexists with a first-order transition, signaled by metastability and by an additional symmetric Liouvillian mode. At stronger pump power the mean-field dynamics loses all stable fixed points and develops limit cycles and chaotic attractors with positive Lyapunov exponent. Quantum trajectories in this regime display chaotic-like motion, enhanced steady-state entropy, and delocalization over many entropic Liouvillian modes. These results establish two-photon-driven optomechanics as a platform where dissipative criticality, symmetry breaking, and quantum signatures of chaos emerge within the same experimentally accessible setting.
\end{abstract}

\maketitle

\section{Introduction}

Cavity optomechanics studies the interaction between the electromagnetic field and mechanical motion mediated by radiation-pressure forces~\cite{Kippenberg2007CavityOptoMechanics,Aspelmeyer2014CavityOptomechanics}.
Over the last two decades, optomechanical systems have emerged as a versatile platform for exploring both fundamental quantum physics and technological applications.
The high degree of controllability achievable in optomechanical platforms has enabled the prediction and observation of a wide variety of nonlinear and quantum phenomena.
These include mechanical ground-state cooling~\cite{Elste2009QuantumNoiseInterference,Schliesser2009ResolvedSidebandCooling,Teufel2011SidebandCoolingMicromechanical,Chan2011LaserCoolingNanomechanical,Whittle2021Approaching}, optomechanically induced transparency~\cite{Fleischhauer2005Electromagnetically,Agarwal2010Electromagnetically,Weis2010OptomechanicallyInducedTransparency}, the generation of nonclassical states of light and motion~\cite{OConnell2010QuantumGroundState,Nunnenkamp2011SinglePhotonOptomechanics,Barzanjeh2021OptomechanicsQuantumTechnologies,Macri2018,Settineri2019,Hauer2023NonlinearSidebandCooling}, squeezing and entanglement generation~\cite{Fabre1994QuantumNoise,SafaviNaeini2013SqueezedLightSilicon,Wollman2015QuantumSqueezingMotion,Pirkkalainen2015Squeezing,Riedinger2018RemoteQuantumEntanglement,OckeloenKorppi2018StabilizedEntanglementMassive,Youssefi2023SqueezedMechanicalOscillator,Palomaki2013EntanglingMechanicalMotion,Riedinger2016NonclassicalCorrelations,Macri2016,Wang2023,Mercurio2025BilateralPhotonEmission}, and quantum transduction, information processing~\cite{Stannigel2011Optomechanical,Pechal2018Superconducting,Wallucks2020AQuantumMemory,DiStefano2019,Russo2023OptomechanicalTwoPhoton}.

Optomechanical systems are also central to precision platforms for gravitational-wave and dark-matter detection~\cite{Abbott2009Observation,Abbott2016Observation,Whittle2021Approaching,Carney2021Mechanical}.
This level of control places optomechanical systems in the regime where the interplay between coherent drives, nonlinearities, and dissipation can be exploited to probe complex and critical non-equilibrium phenomena, such as dissipative chaos and dissipative phase transitions (DPTs).

Chaos is a dynamical manifestation of the underlying complexity of a system.
In classical dynamical systems, chaos is rooted in the exponential sensitivity to initial conditions in the equations of motion~\cite{strogatz_nonlinear_2018}.
In quantum mechanical systems, chaotic behavior is associated with an emergent random matrix structure in the spectrum of the Hamiltonian~\cite{bohigas_characterization_1984,dalessio_quantum_2016,haake_quantum_2018}.
In open quantum systems, the characterization of chaos is more subtle, and several dedicated tools and interpretations have been developed for this purpose~\cite{grobe_quantum_1988,akemann_universal_2019,sa_complex_2020,hamazaki_universality_2020,kawabata_symmetry_2023,villasenor_breakdown_2024}.
Chaotic behavior has been investigated in a variety of driven-dissipative bosonic platforms, with particular focus on circuit and cavity quantum electrodynamics~\cite{dahan_classical_2022,prasad_dissipative_2022,ray_ergodic_2024,peyruchat_landauzener_2025,ferrari_chaotic_2025,kruglikov_chaos_2025,rufo2025quantumsemiclassicalsignaturesdissipative, sa2026talktalkdissipativequantum}, and with growing attention to the performance of quantum devices~\cite{ferrari_dissipative_2025,ferrari2026bitflipssaturationquantum}.

DPTs extend the concept of quantum phase transitions to open quantum systems, whose steady state undergoes a non-analytic change as a control parameter is varied~\cite{minganti_spectral_2018}.
As in equilibrium phase transitions, DPTs can be classified according to the nature of this non-analyticity.
First-order DPTs are characterized by a discontinuous change of the steady state, or of an associated order parameter, and are typically accompanied by metastability and bistability.
Second-order DPTs instead display a continuous order parameter whose derivatives become non-analytic at the critical point.
When the Liouvillian dynamics possesses a symmetry, a second-order DPT can be associated with spontaneous symmetry breaking: the steady state remains symmetric at finite size, while in the thermodynamic limit the system can select one of several symmetry-related states.
DPTs have been widely studied~\cite{kessler_dissipative_2012,Carmichael2015Breakdown,bartolo_exact_2016,biondi_nonequilibrium_2017,Vukics2019Finitesize,minganti_dissipative_2023,mercurio2026floquetdissipativephasetransitions} and experimentally probed in polariton microcavities~\cite{rodriguez_probing_2017,fink_observation_2017} and superconducting circuits~\cite{fitzpatrick_observation_2017,fedorov_photon_2021,chen_quantum_2023,beaulieu_observation_2025}, with applications in, e.g., quantum metrology~\cite{heugel_quantum_2019,garbe_critical_2020,di_candia_critical_2023,alushi_optimality_2024,alushi_collective_2025,beaulieu_criticality-enhanced_2025, mihailescu_critical_2026}.
In cavity optomechanics, DPTs have been theoretically addressed within a classical and Gaussian approximation in single-photon driven architectures~\cite{bibak_dissipative_2023}.
In such systems, however, the coherent single-photon drive explicitly fixes the phase of the cavity field and does not provide the discrete cavity symmetry required for a symmetry-breaking transition.
This leaves open the question of whether optomechanical platforms can host DPTs in which a cavity-field symmetry is spontaneously broken, and how such critical behavior coexists with the nonlinear dynamical instabilities that underlie chaos.
Two-photon driving provides a natural route to address these questions.
Unlike a single-photon drive, a two-photon drive is intrinsically nonlinear and preserves the discrete transformation $\hat{a}\to-\hat{a}$ of the cavity mode.
The corresponding $\mathbb{Z}_2$ symmetry allows the Liouvillian dynamics to split into symmetry sectors and makes possible DPTs associated with spontaneous symmetry breaking.

In this work we investigate DPTs and chaos in a two-photon driven quantum optomechanical system.
First, we characterize this landscape at the mean-field level, deriving steady-state solutions and analytically studying their stability.
We then turn to the full quantum dynamics by analyzing the spectral properties of the Liouvillian superoperator.
We identify the signatures of both a second-order DPT with spontaneous symmetry breaking and a first-order DPT with symmetry breaking, which are further characterized through their Wigner functions and the behavior of individual quantum trajectories.
Finally, we demonstrate the existence of a regime where the classical dynamics acquires a positive Lyapunov exponent~\cite{strogatz_nonlinear_2018}, signaling the onset of classical dissipative chaos.
In this regime, we employ the spectral statistics of quantum trajectories~\cite{ferrari_dissipative_2025} and study the delocalization of the stochastic wavefunction in the Liouvillian spectrum~\cite{richter2025localizationdelocalizationquantumtrajectories} to identify a set of features compatible with dissipative chaos in the presence of classical instability. 

The paper is structured as follows.
In~\cref{sec:model} we introduce the quantum optomechanical setup and discuss its Liouvillian symmetries.
In~\cref{sec: semiclassApp} we derive the semiclassical approximation and the stability map of the system.
In~\cref{sec: DPT} we study the emergence of dissipative phase transitions using the spectral theory of Liouvillians.
In~\cref{sec:chaos} we study the emergence of chaos in the setup from the classical to the full quantum picture.
Finally, in~\cref{sec:conclusion} we draw our conclusions and discuss the outlooks.

\section{Quantum model and symmetries}\label{sec:model}

We consider a two-photon-driven optical resonator (described by the bosonic annihilation operator $\hat{a}$) coupled to a mechanical oscillator (described by the operator $\hat{b}$) \cite{Aspelmeyer2014CavityOptomechanics}.
The optomechanical setup is sketched in~\cref{fig:meanfield sketch}.
In the frame rotating at the pump frequency $\omega_p$ the system is described by the Hamiltonian
\begin{equation}\label{eq:hamiltonian}
    \hat{H} = \Delta\hat{a}^\dagger \hat{a} + \Omega_m\hat{b}^\dagger \hat{b} - g\,\hat{a}^\dagger \hat{a}\,(\hat{b} + \hat{b}^\dagger) + \frac{G}{2}(\hat{a}^2 + \hat{a}^{\dagger 2}) \, .
\end{equation}
Here, $\Delta = \omega_c-\omega_p/2$ is the cavity-pump detuning, $\Omega_m$ denotes the mechanical resonator frequency, $g$ quantifies the optomechanical coupling and $G$ (assumed real without loss of generality) sets the strength of the two-photon drive.
The system's dynamics is governed by the Lindblad master equation~\cite{BreuerPetruccione2002}
\begin{equation}\label{eq:lindblad}
\begin{split}
    \frac{\rmd\hat{\rho}}{\rmd t} =& \ \mathcal{L}\hat{\rho} = -i[\hat{H},\hat{\rho}] + \kappa_1\, \mathcal{D}[\hat{a}]\hat{\rho}  \\
    &+ \kappa_2\, \mathcal{D}[\hat{a}^2]\hat{\rho} + \Gamma_m\, \mathcal{D}[\hat{b}]\hat{\rho} \, ,
\end{split}
\end{equation}
where the Liouvillian $\mathcal{L}$ is the generator of the dissipative time evolution. The parameters $\kappa_1$, $\kappa_2$, and $\Gamma_m$ denote the single-photon, two-photon, and mechanical decay rates, respectively. The Lindblad dissipator acting on an arbitrary jump operator $\hat{L}$ is defined as $\mathcal{D}[\hat{L}]\hat{\rho} = \hat{L}\hat{\rho} \hat{L}^\dagger - \{\hat{L}^\dagger \hat{L},\,\hat{\rho}\}/2$.

The model exhibits a discrete $\mathbb{Z}_2$ weak symmetry associated with the optical mode, corresponding to the invariance under the transformation $\hat{a} \rightarrow -\hat{a}$. The associated symmetry superoperator reads $\mathcal{Z}_2 = e^{i\pi \hat{a}^\dagger \hat{a}} \, \cdot \, e^{-i\pi \hat{a}^\dagger \hat{a}}.$
This implies that the Liouvillian commutes with $\mathcal{Z}_2$, $[\mathcal{L}, \mathcal{Z}_2] = 0$, which allows one to decompose $\mathcal{L}$ into independent symmetry sectors labeled by the eigenvalues of $\mathcal{Z}_2$, namely $z_{\pm 1} = \pm 1$:
\begin{equation}\label{eq:block_diagonalization}
\mathcal{L} =
\begin{pmatrix}
\mathcal{L}_{1} & 0 \\
0 & \mathcal{L}_{-1}
\end{pmatrix} \, ,
\end{equation}
where block $\mathcal{L}_{1}$ contains density-matrix elements connecting states within the $+1$ parity sector, whereas $\mathcal{L}_{-1}$ connects states within the $-1$ sector.
The steady-state density matrix defined via $\mathcal{L}\hat{\rho}_{\rm ss}=0$ satisfies $\mathcal{Z}_2 \rho_{\mathrm{ss}} = \rho_{\mathrm{ss}}$, and therefore belongs to the symmetry sector with eigenvalue $1$.

\section{Semiclassical approximation}
\label{sec: semiclassApp}

\begin{figure*}[t]
    \centering
    \includegraphics{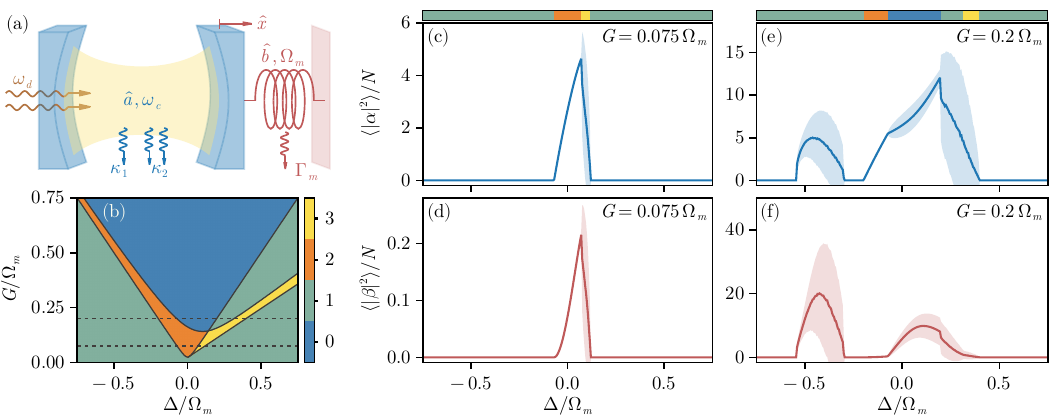}
    \caption{
    Mean-field analysis of the two-photon optomechanical system.
    (a) Schematic representation of the setup. An optical cavity (in blue) is coupled to a mechanical oscillator (in red). The cavity is driven by a two-photon pump and is subject to both single- and two-photon losses, with rates $\kappa_1$ and $\kappa_2$, respectively. The mechanical oscillator has frequency $\Omega_m$ and damping rate $\Gamma_m$.
    (b) Stability diagram in the $(\Delta/\Omega_m,G/\Omega_m)$ plane. The colors indicate the number of physical and stable mean-field solutions, while the dashed horizontal lines mark the two cuts used in the mean-field simulations.
    (c),(d) Mean photon (blue) and phonon (red) occupations along the low-pump cut $G=0.075\,\Omega_m$. In this regime, the system displays a continuous transition at negative detuning and a discontinuous transition in the bistable region at positive detuning.
    (e),(f) Mean photon (blue) and phonon (red) occupations along the high-pump cut $G=0.2\,\Omega_m$. The increased variance around $\Delta/\Omega_m\simeq 0$ is associated with oscillatory dynamics rather than bistability.
    All solid lines in panels (c),(d),(e),(f) represent averages over $5000$ trajectories, while shaded regions indicate the variance computed from the asymptotic time window, as described in the main text. The color bars above panels (c),(e) reproduce the corresponding stability regions from panel (b).
    All other parameters are fixed as follows: $\tilde{g}=0.1\,\Omega_m$, $\kappa_1=0.05\,\Omega_m$, $\tilde{\kappa}_2=0.01\,\Omega_m$, and $\Gamma_m=0.025\,\Omega_m$.
    \label{fig:meanfield}
    }
    {\phantomsubcaptionlabel{fig:meanfield sketch}}
    {\phantomsubcaptionlabel{fig:meanfield stability}}
    {\phantomsubcaptionlabel{fig:meanfield opt weak}}
    {\phantomsubcaptionlabel{fig:meanfield mech weak}}
    {\phantomsubcaptionlabel{fig:meanfield opt strong}}
    {\phantomsubcaptionlabel{fig:meanfield mech strong}}
\end{figure*}

The quantum model in Eq.~\eqref{eq:hamiltonian} admits a mean-field description in terms of complex field coherence $\alpha(t) = \operatorname{Tr}[\hat{\rho}(t)\hat{a}]$ and $\beta(t) = \operatorname{Tr}[\hat{\rho}(t)\hat{b}]$ (the so-called coherent state or Gross-Pitaevskii approximation~\cite{casteels_quantum_2017, debnath_nonequilibrium_2017, giraldo_driven-dissipative_2020}).
The semiclassical equations of motion read
\begin{align}
    \label{eq:eoms_alpha}
    \dot\alpha &= -\qty(i\Delta + \frac{\kappa_1}{2})\alpha + 2i g \textrm{Re}(\beta)\alpha - \kappa_2 |\alpha|^2 \alpha - i G \alpha^* \, ,\\
    \label{eq:eoms_beta}
    \dot\beta  &= -\qty(\rmi\Omega_m + \frac{\Gamma_m}{2})\beta + i g |\alpha|^2 \, .
\end{align}
The thermodynamic limit of \cref{eq:eoms_alpha,eq:eoms_beta} coincides with diverging populations in both oscillators, $\alpha\to\sqrt{N}\,\alpha$ and $\beta\to\sqrt{N}\,\beta$, and the scaling $g\to g/\sqrt{N}$, $\kappa_2\to\kappa_2/N$.
This guarantees that the structure of the equations of motion does not change with $N$.

\subsection{\label{Sec: Physicality}Semiclassical solutions and their stability}

The steady-state solutions are obtained by imposing $\dot{\alpha}=0$ and $\dot{\beta}=0$ in \cref{eq:eoms_alpha,eq:eoms_beta}. 
The second equation leads to the following condition on the mechanical field
\begin{equation}
    \beta_s+\beta_s^*
    =
    \frac{U_{\rm eff}}{g}|\alpha|^2 \, ,
    \quad
    U_{\rm eff} =
    \frac{2g^2\Omega_m}
    {\Omega_m^2+\left(\frac{\Gamma_m}{2}\right)^2} \, ,
    \label{eq: real beta}
\end{equation}
which reduces the problem to the effective nonlinear equation for the optical field
\begin{equation}\label{eq:simil_kerr}
     \left(-i\Delta - \frac{\kappa_1}{2}\right)\alpha + i\,U_{\rm eff}\,\lvert\alpha\rvert^2\alpha - \kappa_2 |\alpha|^2 \alpha - i G \alpha^*=0 \, .
\end{equation}
We write $\alpha=\sqrt{n}_ae^{i\phi_a}$, with $n_a$ and $\phi_a$ respectively the amplitude and phase of the photonic field. 
We find three stationary solutions for the field amplitude: the trivial solution $n_a=0$ and two nonzero solutions $n^{(\pm)}_a$,
\begin{equation}\label{eq: n_pm}
\begin{split}
    &n^{(\pm)}_a =
    \frac{
    -\kappa_1\kappa_2/2+U_{\rm eff}\Delta}{U_{\rm eff}^2+\kappa_2^2}
    \\&\pm
    \frac{\sqrt{
    (U_{\rm eff}^2+\kappa_2^2)G^2
    -
    \left(\kappa_2\Delta+U_{\rm eff}\kappa_1/2\right)^2
    }}
    {U_{\rm eff}^2+\kappa_2^2} \, .
\end{split}
\end{equation}
For each $n^{(\pm)}_a$, the equation for $\phi_a$ has two solutions separated by $\pi$, reflecting the $\mathbb{Z}_2$ symmetry of the system. 
Thus, the mean-field equations admit five stationary solutions: a vacuum-like state and two pairs of symmetry-related finite-amplitude states.
These are valid physical states only when $n_a^{(\pm)}$ are real and positive. 
A complete derivation of the physicality conditions and the corresponding boundaries in the $(\Delta/\Omega_m,\,G/\Omega_m)$ plane is discussed in the Appendix~\ref{app: physicality and stability}.

The stationary solutions identified above are not necessarily dynamically stable. To determine their stability, we employ the Routh-Hurwitz criterion~\cite{DeJesus1987RouthHurwitz}, which is further discussed in Appendix~\ref{app: physicality and stability}.
This criterion predicts that the vacuum solution becomes unstable at the parametric threshold
\begin{equation}
    G_\mathrm{thr}
    =
    \sqrt{
    \Delta^2+\left(\frac{\kappa_1}{2}\right)^2
    } \, .
    \label{eq: threshold}
\end{equation}
The Routh-Hurwitz analysis applied to the nonzero solutions determines the full stability diagram, which is reported in~\cref{fig:meanfield stability}.
The structure of the phase diagram resembles that of a standard two-photon driven Kerr resonator~\cite{sepulcre_analytical_2026}, with new stability regions originating from the purely optomechanical coupling. 
The stability regions with one, two and three solutions, resemble the properties of the standard two-photon driven Kerr resonator.
However,~\cref{fig:meanfield stability} also displays a large region with \textit{zero} stable mean-field solutions.
As we will show in~\cref{sec:chaos}, this is connected with the emergence of limit cycles and chaos~\cite{strogatz_nonlinear_2018} in the system, from the classical to the full quantum regime.

Next, we investigate the different phases appearing in \cref{fig:meanfield stability} by numerically solving \cref{eq:eoms_alpha,eq:eoms_beta}.
We select two values of the drive amplitude, corresponding to a weak pump ($G/\Omega_m=0.075$) and a stronger one ($G/\Omega_m=0.2$), and vary $\Delta$.
For each value of $\Delta/\Omega_m\in[-0.75,0.75]$, 
we sample $5000$ initial fields $\alpha(0)$ and $\beta(0)$ from the complex Gaussian distribution $\mathcal{N}(0, 2)$. 
Each trajectory is evolved up to a final time $t\gg\textrm{max}\{1/\kappa_1,\,1/\kappa_2,\,1/\Omega_m\}$.
In~\cref{fig:meanfield opt weak,fig:meanfield mech weak} we report the populations averaged over the final $10\%$ of the simulation time (taken as the asymptotic window) across the $5000$ independent trajectories for $G/\Omega_m=0.075$.
Similarly,~\cref{fig:meanfield opt strong,fig:meanfield mech strong} show the corresponding populations for the stronger drive $G/\Omega_m=0.2$.

At weak pumping, the mean-field photon and phonon numbers show a continuous transition from the vacuum to a populated state at small negative detunings. 
At small positive detunings, where the mean-field equations admit bistable solutions, $\langle|\alpha|^2\rangle$ and $\langle|\beta|^2\rangle$ display a discontinuous jump.
The shaded region around the curves represents the steady-state variance of the populations, obtained directly by computing the variance over the final $10\%$ of the trajectories across all $5000$ random initial conditions.
As expected, the variance is large only in presence of optical bistability.

In the strong pumping regime, the system enters a phase with no stable fixed points.
We observe that a significantly broader range of detunings now correspond to an increased population variance. This fluctuation is not merely linked to bistability, but signals the emergence of oscillatory dynamics, either regular (limit cycles) or chaotic (strange attractors).
Moreover, we observe that around $\Delta/\Omega_m=-0.5$ a secondary peak appears, characterized by $\langle|\beta|^2\rangle>\langle|\alpha|^2\rangle$, which is also accompanied by an enhanced mean-field variance.
We discuss the features of this strong-pump regime in greater depth in~\cref{sec:chaos}

Throughout the main text, we set the strength of the optomechanical coupling to $g/\Omega_m=0.1$ and the mechanical losses to $\Gamma_m/\Omega_m=0.025$.
In~\cref{sec:exp_parameters} we study in detail~\cref{eq:hamiltonian,eq:lindblad} with typical experimental parameters~\cite{Youssefi2023SqueezedMechanicalOscillator}.  

\section{Dissipative phase transitions}\label{sec: DPT}

The semiclassical analysis identifies the different operating regimes of the system, and in particular characterizes the features of the steady-state mean-field solutions.
The behavior of the semiclassical amplitudes reveals hallmarks of dissipative phase transitions~\cite{minganti_spectral_2018}, such as the continuous yet sharp transition from the vacuum to a populated state at negative $\Delta/\Omega_m$, alongside a discontinuous jump at positive $\Delta/\Omega_m$.
To characterize the emergence of DPTs in the quantum regime, we analyze the steady-state density matrix alongside the low-lying spectrum of the Liouvillian~\cite{minganti_spectral_2018}.
These defining spectral features are obtained numerically via the Arnoldi-Lindblad procedure~\cite{Minganti2022arnoldilindbladtime}.
All the numerical simulations presented in this work have been performed with the \texttt{QuantumToolbox.jl} package~\cite{QuantumToolbox.jl2025} available in Julia. 

The presence of a weak $\mathbb{Z}_2$ symmetry in the system gives rise to first- and second-order DPTs.
Specifically, the setup we consider hosts both phenomena: a second-order DPT for $\Delta/\Omega_m < 0$ and a first-order DPT for $\Delta/\Omega_m > 0$.

\subsection{Second-order DPT}
We begin our analysis of the second-order DPT by studying a cut of the phase diagram with fixed detuning $\Delta/\Omega_m = -0.065$.
We vary the two-photon drive amplitude $G/\Omega_m$ and the parameter $N$ quantifying the distance from the thermodynamic limit. \cref{fig:2ndOrder photon} shows the rescaled photon and phonon numbers in the steady state, $\langle \hat{a}^\dagger \hat{a} \rangle/N$ (main panel) and $\langle \hat{b}^\dagger \hat{b} \rangle/N$ (inset), as a function of $G/\Omega_m$.
The photon number becomes sharper with $N$, and
both the observables change smoothly across the transition.
This is in agreement with the continuous behavior expected for a second-order DPT.

The spectral signature of the transition is encoded in the behavior of the Liouvillian gap $-\mathrm{Re}[\lambda_1/\Omega_m]$~\cite{minganti_spectral_2018}, which we plot in~\cref{fig:2ndOrder gap}.
As $N$ increases, the gap closes once the critical point of the second-order DPT is crossed, i.e., throughout the symmetry-broken phase. 
Here, $\lambda_1\in\mathbb{R}\to0$ and the associated Hermitian and traceless eigenoperator $\hat{\rho}_1$ belongs to the odd parity sector of the $\mathbb{Z}_2$ symmetry superoperator, such that $\mathcal{Z}_2 \hat{\rho}_1 = -\hat{\rho}_1$~\cite{minganti_spectral_2018}.
Since $\operatorname{Tr}(\hat{\rho}_1)=0$, the eigenoperator $\hat{\rho}_1$ can be decomposed into the form $\hat{\rho}_1 \propto \hat{\rho}_+ - \hat{\rho}_-$, where $\hat{\rho}_\pm$ are physical density matrices. 
Furthermore, these density matrices are mapped onto one another by the $\mathbb{Z}_2$ transformation, such that $\mathcal{Z}_2\hat{\rho}_+ = \hat{\rho}_-$.
In the symmetry-broken phase, the steady state $\hat{\rho}_{\mathrm{ss}}$ is expected to inherit the structure of $\hat{\rho}_1$ while preserving unit trace.
These conditions suggest the Ansatz~\cite{minganti_spectral_2018}
\begin{equation}
    \hat{\rho}_{\mathrm{ss}}\simeq\frac{\hat{\rho}_+ + \hat{\rho}_-}{2} = \hat{\xi}.
\end{equation}
We plot the fidelity between $\hat{\rho}_{\rm ss}$ and $\hat{\xi}$, $\mathcal{F}(\hat{\rho}_{\rm ss}, \hat{\xi}) = \operatorname{Tr}^2[(\sqrt{\hat{\rho}_{\rm ss}}\,\hat{\xi}\,\sqrt{\hat{\rho}_{\rm ss}})^{1/2}]$ in the inset of~\cref{fig:2ndOrder gap}.
As the system approaches the thermodynamic limit, $1-\mathcal{F}$ strictly vanishes throughout the symmetry-broken phase.

\begin{figure}[b]
    \centering
    \includegraphics{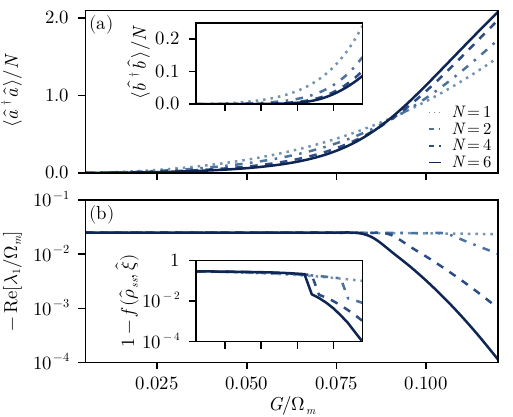}
    \caption{
    Emergence of a second-order dissipative phase transition as a function of the pump strength $G/\Omega_m$ for increasing scaling parameter $N$.
    (a) Rescaled photon occupation $\langle \hat{a}^\dagger \hat{a} \rangle/N$. The smooth increase of the occupation across the critical point is consistent with a continuous transition. The inset shows the corresponding rescaled phonon occupation $\langle \hat{b}^\dagger \hat{b} \rangle/N$.
    (b) Rescaled Liouvillian gap $-\mathrm{Re}[\lambda_1/\Omega_m]$. As $N$ increases, the gap decreases and remains close to zero over an extended region, signaling the emergence of the symmetry-broken phase. The inset shows the error $1-f(\rho_{\mathrm{ss}},\xi)$, where $f$ is the fidelity between the steady state and its reconstructed form $\xi=(\rho_+ + \rho_-)/2$. The decrease of this error indicates that the spectral reconstruction becomes more accurate as the thermodynamic limit is approached.
    Other parameters are set as in \cref{fig:meanfield}.
    \label{fig:2ndOrder}
    }
    {\phantomsubcaptionlabel{fig:2ndOrder photon}}
    {\phantomsubcaptionlabel{fig:2ndOrder gap}}
\end{figure}

\begin{figure}[t]
    \centering
    \includegraphics{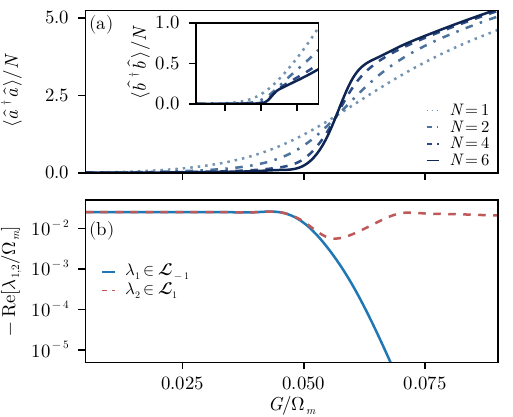}
    \caption{
    Emergence of a first-order dissipative phase transition with symmetry breaking as a function of the pump strength $G/\Omega_m$ for increasing $N$.
    (a) Rescaled photon occupation $\langle \hat{a}^\dagger \hat{a} \rangle/N$, displaying a sharp jump that becomes more abrupt as $N$ increases. The inset shows the corresponding rescaled phonon occupation $\langle \hat{b}^\dagger \hat{b} \rangle/N$.
    (b) Real parts of the two Liouvillian eigenvalues with smallest modulus, $-\mathrm{Re}[\lambda_{1,2}/\Omega_m]$, for $N=5$. The eigenvalue $\lambda_1\in\mathcal{L}_{-1}$ (blue solid line) becomes small throughout the symmetry-broken region, while $\lambda_2\in\mathcal{L}_{1}$ (red dashed line) becomes small only close to the critical point. This indicates the coexistence of metastable states and the discontinuous nature of the transition.
    Other parameters are the same as in \cref{fig:meanfield}.
    \label{fig:1stOrder}}
    {\phantomsubcaptionlabel{fig:1stOrder population}}
   {\phantomsubcaptionlabel{fig:1stOrder gap}}
\end{figure}

\subsection{First-order DPT with symmetry breaking}

\begin{figure*}[t!]
    \centering
    \includegraphics{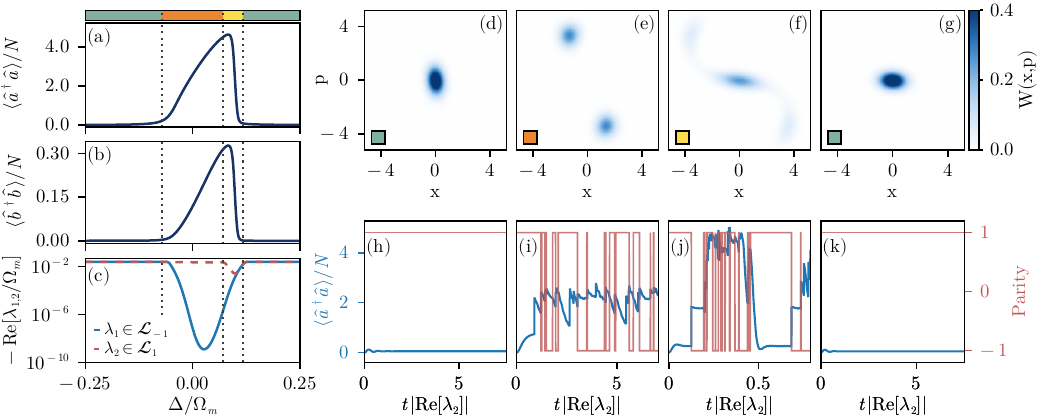}
    \caption{
    Steady-state and phase-space characterization as a function of the detuning $\Delta/\Omega_m$ at fixed pump strength $G=0.075\,\Omega_m$ and system size $N=5$.
    (a),(b) Rescaled photon and phonon occupations, $\langle \hat{a}^\dagger \hat{a} \rangle/N$ and $\langle \hat{b}^\dagger \hat{b} \rangle/N$. The occupations show a smooth onset at negative detuning and a sharp drop at positive detuning.
    (c) Real parts of the two Liouvillian eigenvalues with smallest modulus, $-\mathrm{Re}[\lambda_{1,2}/\Omega_m]$. The eigenvalue $\lambda_1\in\mathcal{L}_{-1}$ remains small over the symmetry-broken region, while $\lambda_2\in\mathcal{L}_{1}$ becomes small only close to the positive-detuning transition, signaling the first-order transition.
    (d)--(g) Wigner functions of the optical mode for representative detunings chosen inside the regions separated by the vertical dashed lines in panels (a)--(c). The values are $\Delta/\Omega_m=-0.15,\,0.0,\,0.1,\,0.2$. The colored markers identify the corresponding regions in the stability map.
    (h)--(k) Single quantum trajectories for the same detunings, showing the rescaled photon number $\langle \hat{a}^\dagger \hat{a} \rangle/N$ in blue and the parity in red. The horizontal axis is expressed in units of $t|\mathrm{Re}[\lambda_{2}]|$, where $\lambda_{2}$ is the lowest nonzero eigenvalue in the symmetric Liouvillian sector, which sets the metastable switching time scale.
    In the symmetric regions, the Wigner function has a single peak and the trajectories remain close to the low-occupation state with fixed parity. In the symmetry-broken region, the Wigner function becomes bimodal and the trajectories switch between the two symmetry-related states. In the bistable region, the Wigner function shows the coexistence of a central peak and two symmetry-broken lobes, while the trajectory alternates between low- and high-occupation regimes.
    Other parameters are set as in \cref{fig:meanfield}.
    \label{fig:delta_scan}}
    {\phantomsubcaptionlabel{fig:delta_scan photon}}
    {\phantomsubcaptionlabel{fig:delta_scan phonon}}
    {\phantomsubcaptionlabel{fig:delta_scan gap}}
    {\phantomsubcaptionlabel{fig:delta_scan wigner 1}}
    {\phantomsubcaptionlabel{fig:delta_scan wigner 2}}
    {\phantomsubcaptionlabel{fig:delta_scan wigner 3}}
    {\phantomsubcaptionlabel{fig:delta_scan wigner 4}}
    {\phantomsubcaptionlabel{fig:delta_scan traj 1}}
    {\phantomsubcaptionlabel{fig:delta_scan traj 2}}
    {\phantomsubcaptionlabel{fig:delta_scan traj 3}}
    {\phantomsubcaptionlabel{fig:delta_scan traj 4}}
\end{figure*}

We now study the first-order DPT with symmetry breaking by fixing $\Delta/\Omega_m=0.065$ and varying $G/\Omega_m$ and $N$.
\cref{fig:1stOrder population} shows the populations $\langle\hat{a}^\dagger\hat{a}\rangle/N$ (main panel) and $\langle\hat{b}^\dagger\hat{b}\rangle/N$ (inset) as a function of $G/\Omega_m$.
In contrast to the second-order DPT, the photonic and phononic modes become abruptly populated around the critical point, with this onset becoming increasingly sharp as $N$ grows.
This is a typical feature of first-order transitions~\cite{bartolo_exact_2016, minganti_spectral_2018} and is reminiscent of the sudden jump in the semiclassical photon and phonon numbers reported in~\cref{fig:meanfield opt weak,fig:meanfield mech weak}.

At the spectral level, several differences arise with respect to the previous case.
For the second-order DPT, a single eigenvalue $\lambda_1$ belonging to the \textit{antisymmetric} Liouvillian sector $\mathcal{L}_{-1}$ (cf.~\cref{eq:block_diagonalization}) approaches zero.
For the first-order DPT with symmetry breaking, a second eigenvalue $\lambda_2$, belonging to the \textit{symmetric} Liouvillian sector $\mathcal{L}_{1}$ collapses to 0.
While $\lambda_1\to0$ in the whole symmetry-broken phase, $\lambda_2\to0$ only in the vicinity of the critical point~\cite{minganti_spectral_2018}.
This second Liouvillian mode is associated with the discontinuous change of the steady state and is therefore the spectral fingerprint of the first-order transition.
We plot $\textrm{Re}[\lambda_1/\Omega_m]$ and $\textrm{Re}[\lambda_2/\Omega_m]$ as a function of $G/\Omega_m$ and for $N = 5$ in~\cref{fig:1stOrder gap}.
The behavior of $\lambda_1$ identifies the symmetry-broken region, while the closing of $\lambda_2$ near the critical point signals the coexistence of metastable states and determines the discontinuous switching between the two phases.

\subsection{Phase-space characterization}
\label{sec:phasespace}
To further characterize the dynamics across both DPTs in the two-photon-driven dissipative optomechanical system, we analyze its phase-space dynamics and quantum trajectories.
We fix the two-photon pump strength to $G=0.075\,\Omega_m$ and vary the detuning $\Delta/\Omega_m$, thus crossing both transitions discussed above and allowing a direct comparison with the mean-field results.

The steady-state photon and phonon numbers are shown in~\cref{fig:delta_scan photon,fig:delta_scan phonon} respectively. 
For negative detuning, the photon and phonon occupations increase smoothly, consistently with a continuous transition. 
For positive detuning, instead, the sharp drop of the occupations signals the first-order transition. 
The Liouvillian spectrum (cf.~\cref{fig:delta_scan gap}) confirms this interpretation: in the negative-detuning region, a low-lying eigenvalue in the antisymmetric sector $\mathcal{L}_{-1}$ remains close to zero in the symmetry-broken phase, whereas, near the first-order DPT, an additional eigenvalue in the symmetric sector $\mathcal{L}_{1}$ becomes small only close to the critical point.

To visualize the behavior of the system as $\Delta/\Omega_m$ is varied, in~\cref{fig:delta_scan wigner 1,fig:delta_scan wigner 2,fig:delta_scan wigner 3,fig:delta_scan wigner 4} we plot the Wigner functions of $\hat{\rho}_{\textrm{ss}, a} = \operatorname{Tr}_b[\hat{\rho}_{\rm ss}]$ associated with the optical mode, defined as~\cite{polkovnikov_phase_2010}
\begin{equation}
    W_a(x,p)
    =
    \frac{1}{2\pi}
    \int_{-\infty}^{\infty}
    dy\,
    e^{-ipy}
    \left\langle x+\frac{y}{2}\right|
    \hat{\rho}_{\textrm{ss}, a}
    \left|x-\frac{y}{2}\right\rangle,
\end{equation}
where $x=\sqrt{2}\textrm{Re}(\alpha)$ and $p=\sqrt{2}\textrm{Im}(\alpha)$.
The steady-state Wigner functions for the system under consideration are always positive, due to the single-photon and phonon losses~\cite{bartolo_exact_2016}, and correspond to a probability density over phase space representing the reduced density matrix $\hat{\rho}_{\textrm{ss}, a}$.
Furthermore, in~\cref{fig:delta_scan traj 1,fig:delta_scan traj 2,fig:delta_scan traj 3,fig:delta_scan traj 4}, we examine the dynamics of single quantum trajectories.
These correspond to individual realizations of stochastic wave functions, whose ensemble average exactly recovers the density matrix evolution governed by the Lindblad master equation \eqref{eq:lindblad}~\cite{dalibard_wave-function_1992, molmer_monte_1993, daley_quantum_2014}.
We compute the photon number $\bra{\Psi(t)}\hat{a}^\dagger\hat{a}\ket{\Psi(t)}/N$ and the expectation value of the parity operator $\bra{\Psi(t)}\hat{P}\ket{\Psi(t)}$ with $\hat{P}=e^{i\pi\hat{a}^\dagger\hat{a}}$ over a single quantum trajectory $\ket{\Psi(t)}$.
We refer the reader to Appendix~\ref{app:quantum_trajectories} for a more detailed discussion on quantum trajectories.

In the regions characterized by a unique semiclassically stable fixed point, either before the second-order DPT (\cref{fig:delta_scan wigner 1,fig:delta_scan traj 1}), or after the first-order DPT (\cref{fig:delta_scan wigner 4,fig:delta_scan traj 4}), the Wigner function is a squeezed vacuum state and the trajectories show small photon-number fluctuations with fixed parity. In the symmetry-broken region (\cref{fig:delta_scan wigner 2,fig:delta_scan traj 2}), the Wigner function becomes bimodal and the trajectories switch between the two symmetry-related configurations, with parity alternating between $\pm 1$. In the bistable region associated with the first-order transition (\cref{fig:delta_scan wigner 3,fig:delta_scan traj 3}), the Wigner function shows the coexistence of a central peak and two symmetry-broken lobes. The corresponding quantum trajectory oscillates between the squeezed vacuum state and a high-occupation symmetry-broken state, providing direct evidence of optical bistability~\cite{drummond_quantum_1980, drummond_quantum_1981}.

\begin{figure*}[t]
    \centering
    \includegraphics{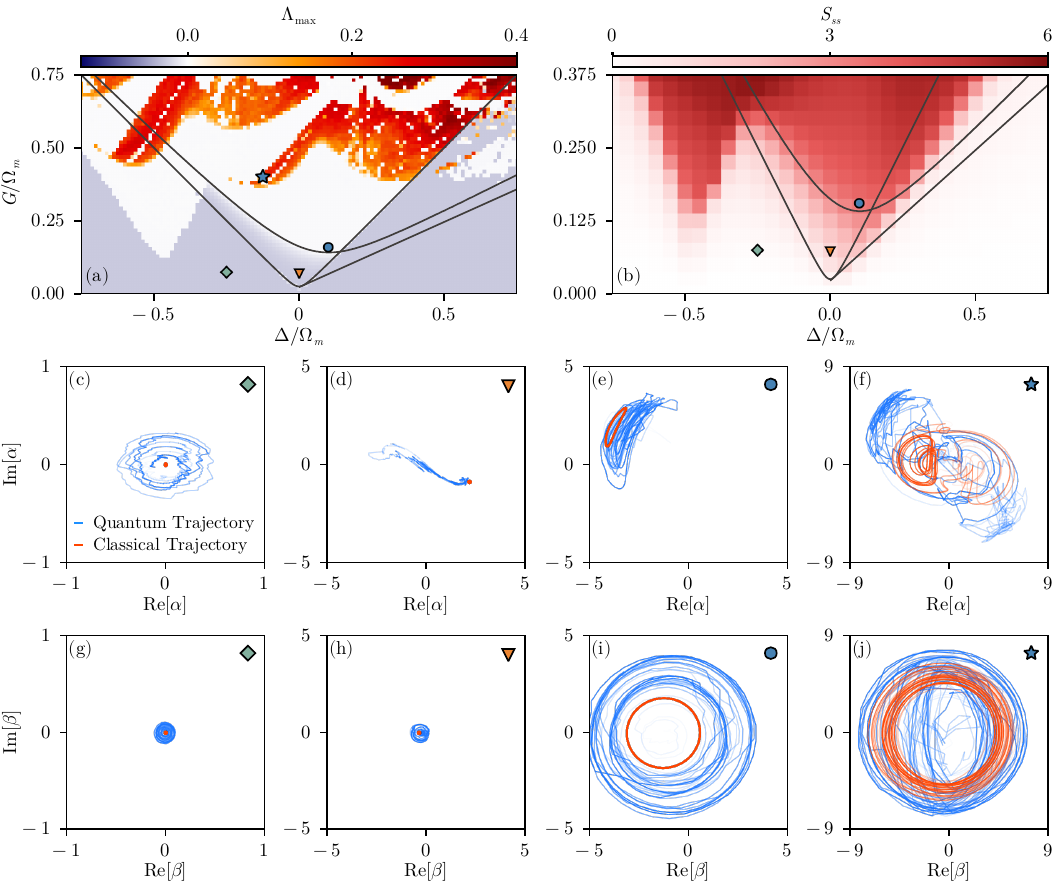}
\caption{
    Classical and quantum characterization of the high-pump dynamical regime.
    (a) Map of the largest Lyapunov exponent $\Lambda_{\mathrm{max}}$ obtained from the mean-field equations. Positive values of $\Lambda_{\mathrm{max}}$ identify classically chaotic dynamics, while regions with $\Lambda_{\mathrm{max}}\simeq 0$ correspond to regular limit cycles. These time-dependent regimes appear inside the unstable region predicted by the mean-field stability diagram of~\cref{fig:meanfield stability}. The side peak observed in the mean-field simulations of~\cref{fig:meanfield opt strong,fig:meanfield mech strong} is also visible in this map. On the right side of the diagram, a region where the stable empty solution coexists with a limit-cycle or chaotic attractor is observed.
    (b) Von Neumann entropy of the steady state, $S_{\mathrm{ss}}=-\mathrm{Tr}[\rho_{\mathrm{ss}}\log\rho_{\mathrm{ss}}]$, showing enhanced mixing in the same high-pump region where the classical dynamics becomes time dependent.
    (c)--(f) Optical phase-space trajectories and (g)--(j) mechanical phase-space trajectories for the representative points marked in panels (a),(b). Blue curves correspond to single stochastic quantum trajectories, while red curves show the corresponding classical mean-field trajectories. The marked points correspond to: diamond (c),(g), $\Delta/\Omega_m=-0.25$, $G/\Omega_m=0.075$; triangle (d),(h), $\Delta/\Omega_m=0$, $G/\Omega_m=0.075$; circle (e),(i), $\Delta/\Omega_m=0.1$, $G/\Omega_m=0.15$; star (f),(j), $\Delta/\Omega_m=-0.125$, $G/\Omega_m=0.4$.
    At low pump strength, panels (c),(g) and (d),(h), both dynamics remain localized close to fixed points. In the time-dependent regime, panels (e),(i), the classical dynamics forms a regular limit cycle, while the quantum trajectory explores a broader region of phase space. At stronger pump, panels (f),(j), both trajectories become chaotic, with the quantum trajectory covering a wider region due to dissipative quantum fluctuations.
    All other parameters are the same as in \cref{fig:meanfield} and $N=1$.
    \label{fig:chaos_map}}
    {\phantomsubcaptionlabel{fig:chaos_map lyapunov}}
    {\phantomsubcaptionlabel{fig:chaos_map entropy}}
    {\phantomsubcaptionlabel{fig:chaos_map opt 1}}
    {\phantomsubcaptionlabel{fig:chaos_map opt 2}}
    {\phantomsubcaptionlabel{fig:chaos_map opt 3}}
    {\phantomsubcaptionlabel{fig:chaos_map opt 4}}
    {\phantomsubcaptionlabel{fig:chaos_map mech 1}}
    {\phantomsubcaptionlabel{fig:chaos_map mech 2}}
    {\phantomsubcaptionlabel{fig:chaos_map mech 3}}
    {\phantomsubcaptionlabel{fig:chaos_map mech 4}}
\end{figure*}

\section{Dissipative chaos}\label{sec:chaos}

Our analysis has so far focused on moderate pump strengths compared to the optomechanical energy, where the semiclassical dynamics is characterized by a nonzero number of stable fixed points (cf.~\cref{fig:meanfield stability}).
In this regime, we extensively characterized the different semiclassical phases at the quantum level using the spectral theory of dissipative phase transitions~\cite{minganti_spectral_2018}.
We now turn our attention to the region characterized by zero stable fixed points at the mean-field level, marked in blue in~\cref{fig:meanfield stability}.
This regime is defined by the emergence of limit cycles and chaotic attractors. 
We first characterize these features in the classical limit before probing the full quantum case for signatures of dissipative chaos.

\subsection{Classical chaotic dynamics}\label{sec:classical_chaos}
The semiclassical equations in \cref{eq:eoms_alpha,eq:eoms_beta} are a four-dimensional autonomous nonlinear system for the variables
\begin{equation}
    \mathbf{x}
    =
    \big(
    \mathrm{Re}[\alpha],\,
    \mathrm{Im}[\alpha],\,
    \mathrm{Re}[\beta],\,
    \mathrm{Im}[\beta]
    \big).
\end{equation}
Consequently, Poincarè-Bendixson theorem does not apply~\cite{strogatz_nonlinear_2018} and the system can display chaotic classical motion.
Classical chaos is defined through an exponential sensitivity to initial conditions. 
Given two nearby trajectories separated by an infinitesimal perturbation $\delta\mathbf{x}(t)$, the rate of exponential divergence is characterized by the Lyapunov exponent \cite{strogatz_nonlinear_2018},
\begin{equation}
    |\delta \mathbf{x}(t)|
    \sim
    |\delta \mathbf{x}(0)| e^{\Lambda t}.
\end{equation}
The largest Lyapunov exponent $\Lambda_{\mathrm{max}}$ distinguishes the different dynamical regimes: $\Lambda_{\mathrm{max}}>0$ signals chaotic dynamics, $\Lambda_{\mathrm{max}}\simeq 0$ corresponds to limit cycles, and $\Lambda_{\mathrm{max}}<0$ indicates relaxation to a stable fixed point.
To calculate $\Lambda_{\rm max}$ from the mean-field equations of motion, we used a method based on the evolution and periodic renormalization of an infinitesimal perturbation \cite{Wolf1985Lyapunov}.

In \cref{fig:chaos_map lyapunov} we plot $\Lambda_{\mathrm{max}}$ as a function of $\Delta/\Omega_m$ and $G/\Omega_m$.
The black lines reproduce the stability map reported in \cref{fig:meanfield stability}.
We observe that the region with zero stable fixed points corresponds to a broad domain with $\Lambda_{\rm max} = 0$, where smaller subregions with $\Lambda_{\max} > 0$ emerge.
In these regions, limit cycles arising from the classical instability deform into chaotic attractors characterized by strictly positive Lyapunov exponents.
We notice how the region with $\Lambda_{\rm max}\ge 0$ extends beyond the black lines delimiting the zero stable solutions domain in the classical limit.
In particular, we notice an additional area around $\Delta/\Omega_m=-0.5$, which corresponds to the peak around the same detuning value reported in~\cref{fig:meanfield opt strong,fig:meanfield mech strong}, where both the photonic and phononic modes populate and exhibit oscillating trajectories, as it results from the large variance around the averaged values of $\langle|\alpha|^2\rangle$ and $\langle|\beta|^2\rangle$.
This mismatch between~\cref{fig:chaos_map lyapunov} and~\cref{fig:meanfield opt strong,fig:meanfield mech strong} and~\cref{fig:meanfield stability} is exclusively due to the initial conditions chosen for \cref{eq:eoms_alpha,eq:eoms_beta}.
In this region, classical trajectories characterized by $|\alpha(0)|$, $|\beta(0)|\gg0$ are captured by a limit cycle or a chaotic attractor, despite the fact that the vacuum is a stable fixed point, as indicated by~\cref{fig:meanfield stability}.
Since the initial conditions are randomly sampled from the Gaussian distribution $\mathcal{N}(0, 2)$, most of the classical trajectories flow in the limit cycle manifold, thus exhibiting $\Lambda_{\rm max}\ge 0$.

Vice versa, if $|\alpha(0)|$, $|\beta(0)|=\varepsilon$ with $\varepsilon\ll1$, then the region hosting limit cycles and strange attractors coincides with the blue zone in~\cref{fig:meanfield stability}, as we show in Appendix~\ref{app:more_on_classical_chaos}.

\subsection{\label{sec:signatures}Quantum signatures of dissipative chaos}

Signatures of quantum chaos are hidden in the spectral correlations of the time evolution generator~\cite{dalessio_quantum_2016}.
For closed quantum systems, consecutive distances between two Hamiltonian eigenvalues exhibit different statistical behaviors according to the model's integrability. Integrable quantum systems are associated with Poissonian level statistics~\cite{berry_integrable_1977} typical of uncorrelated random variables. Chaotic quantum systems on the other hand display level repulsion described by random-matrix theory~\cite{bohigas_characterization_1984, dalessio_quantum_2016}.
For open quantum systems, the same idea has been applied to the Liouvillian: distances among Liouvillian eigenvalues that follow a 2D Poissonian statistics signal integrability, whereas a level statistics conforming to non-Hermitian random matrix theory is connected to dissipative quantum chaos~\cite{grobe_quantum_1988, akemann_universal_2019, hamazaki_universality_2020, sa_complex_2020, sa2026talktalkdissipativequantum}.

\begin{figure*}[t!]
    \centering
    \includegraphics{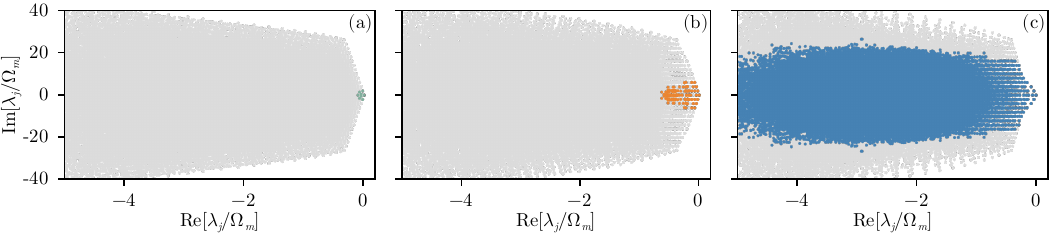}
    \caption{
    Liouvillian spectra and active modes selected by stochastic quantum trajectories.
    (a)--(c) Complex Liouvillian spectra for three different parameter points marked in \cref{fig:chaos_map}: (a) $\Delta/\Omega_m=-0.25$, $G/\Omega_m=0.075$; (b) $\Delta/\Omega_m=0$, $G/\Omega_m=0.075$; (c) $\Delta/\Omega_m=0.1$, $G/\Omega_m=0.15$. Each panel corresponds to a different Liouvillian. Gray points show the full spectrum of the corresponding Liouvillian, while colored points indicate the active modes selected according to the procedure described in Appendix \ref{app:ssqt}. The active modes reported in each panel are those activated by one representative stochastic trajectory at a fixed time $t$. In the regular regimes, only a small localized set of modes is activated, whereas in the limit-cycle regime a broad region of the spectrum is populated.
    All other parameters are the same as in \cref{fig:meanfield} and $N=1$.
    \label{fig:spettri}}
    {\phantomsubcaptionlabel{fig:spettro 1}}
    {\phantomsubcaptionlabel{fig:spettro 2}}
    {\phantomsubcaptionlabel{fig:spettro 3}}
\end{figure*}

\begin{figure}[t!]
    \centering
    \includegraphics{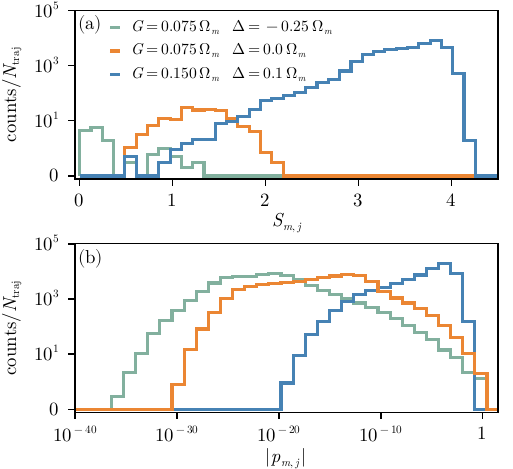}
    \caption{
    Distributions of the mode entropy and of the gauge-independent Liouvillian weights.
    (a) Distribution of the entropy $S_{m,j}$ of the active modes. For each stochastic trajectory, the entropy is computed from the positive component $\rho_{j,+}$ of the Hermitian part of each active mode. The histogram is obtained by collecting the values of $S_{m,j}$ from all trajectories and then binning the full set. The vertical axis reports the counts normalized by the number of trajectories, $\mathrm{counts}/N_{\mathrm{traj}}$, in pseudo-logarithmic scale. In the semiclassical limit-cycle regime, the distribution shifts towards larger values of $S_{m,j}$, indicating higher entropy of the active modes. The increase in the normalized counts also reflects the larger number of active modes selected in this regime.
    (b) Distribution of the gauge-independent weights $|p_{m,j}|$, computed over the full Liouvillian basis. The histogram is obtained by collecting the values of $|p_{m,j}|$ from all trajectories and then binning the full set. The vertical axis again reports $\mathrm{counts}/N_{\mathrm{traj}}$ in pseudo-logarithmic scale. Moving from the low-pump regimes, $G/\Omega_m=0.075$, to the higher-pump regime, $G/\Omega_m=0.15$, many more modes acquire a finite weight and contribute to the quantum trajectory dynamics.
    For each parameter point, 10 stochastic trajectories are used.
    }
    \label{fig:entropie}
    {\phantomsubcaptionlabel{fig:entropie 1}}
    {\phantomsubcaptionlabel{fig:entropie 2}}
\end{figure}

More recently, the idea that spectral statistics can unambiguously identify chaos or integrability in the dynamical phases of open quantum systems has been questioned~\cite{ferrari_chaotic_2025, naves2026levelrepulsionfailsnonnormality}.
We analyze the Liouvillian spectral statistics associated with the driven optomechanical setup in the Appendix~\ref{app: complexspacing}.
We find that the spectral structure of $\mathcal{L}$ does not conform non-Hermitian random matrix theory for any of the parameters configurations we consider, regardless of the classical behavior.
We provide some heuristic arguments for which the spectral statistics could not be reliable in the system we are studying.
While several models studied under the lenses of the spectral criteria, such as spin~\cite{akemann_universal_2019, sa_complex_2020}, fermionic~\cite{kawabata_symmetry_2023}, or bosonic~\cite{ferrari_dissipative_2025} systems, are typically characterized by uniform parameters, our setup displays the clear hierarchy: $\Omega_m\gg g,\Gamma_m>\kappa_2$. 
Thus, the dominant term is a weakly dissipative quantum harmonic oscillator characterized by an equally spaced spectrum.
A possible explanation could rely on a spectrum polluted by a trivial and large contribution coming from $\Omega_m$, while chaotic dynamics is linked to the perturbative interaction terms proportional to $g\ll\Omega_m$ and $\kappa_2$.
We cannot exclude \textit{a priori} that increasing $G/\Omega_m$ (deeper in the classically unstable region and closer to the zones characterized by strange attractors) we would obtain spectral correlations closer to the Ginibre statistics of non-Hermitian random matrices~\cite{akemann_universal_2019}.
However, a reliable diagonalization of $\mathcal{L}$ becomes rapidly unfeasible due to the growth of the photonic and phononic populations with $G$.

To extract possible signatures of chaotic-like behavior, we resort to a number of indirect signatures.
We start by analyzing the structure of the steady-state density matrix.
We then unravel the steady state in stochastic quantum trajectories and we study their properties.
In particular, we employ the spectral statistics of quantum trajectories (SSQT) criterion introduced in Ref.~\cite{ferrari_dissipative_2025} and we argue that the stochastic wave function $\ket{\Psi(t)}$ builds up from an extensive set of highly entropic Liouvillian eigenstates.
As we show below, all these signatures are compatible with dissipative chaotic dynamics.

\subsubsection{Entropy of the steady state}\label{sec:ss_entropy}

First, we analyze the Von Neumann entropy of the steady-state density matrix, defined as
\begin{equation}\label{eq:vn_entropy}
     S_{\rm ss} = -\mathrm{Tr}\left[\hat{\rho}_{\mathrm{ss}}\log\hat{\rho}_{\mathrm{ss}}\right].
\end{equation}
In~\cref{fig:chaos_map entropy} we plot $S_{\rm ss}$ as a function of $\Delta/\Omega_m$ and $G/\Omega_m$, with the black lines reproducing the stability map plotted in~\cref{fig:meanfield stability}.
Notice how the classically unstable region, characterized by limit cycles and chaotic attractors, is characterized by a large entropy in the quantum steady state.
We specifically note the dip around $\Delta/\Omega_m=-0.5$. 
While accessing this feature classically depends strictly on the initial conditions, the quantum system naturally settles into a high-entropy steady state ($S_{\rm ss}\gtrsim4$) regardless of $\hat{\rho}(0)$.
We interpret this finding as evidence that quantum noise inherently destabilizes the vacuum, forcing the stochastic quantum trajectories to asymptotically explore the limit-cycle manifold.
Finally, $S_{\rm ss}\gtrsim\log(2)$ in the region characterized by two classically stable fixed points, as expected by a mixture of two nearly-coherent states (see~\cref{fig:delta_scan wigner 2}).

\subsubsection{Classical and quantum trajectories}\label{sec:quantum_trajs}

Next, we compare the dynamics of individual classical and quantum trajectories. 
Specifically, we consider the classical coherences $\alpha(t)$ and $\beta(t)$ obtained by numerically solving \cref{eq:eoms_alpha,eq:eoms_beta} and the expectation values of the quantum fields $\bra{\Psi(t)}\hat{a}\ket{\Psi(t)}$ and $\bra{\Psi(t)}\hat{b}\ket{\Psi(t)}$ obtained with the procedure detailed in Appendix~\ref{app:quantum_trajectories}.
The results are plotted in~\cref{fig:chaos_map opt 1,fig:chaos_map opt 2,fig:chaos_map opt 3,fig:chaos_map opt 4,fig:chaos_map mech 1,fig:chaos_map mech 2,fig:chaos_map mech 3,fig:chaos_map mech 4} for both the optical and mechanical oscillators.

\cref{fig:chaos_map opt 1,fig:chaos_map mech 1} correspond to the classical region characterized by a single stable fixed point in~\cref{fig:meanfield stability}.
In this case, the classical and quantum trajectories for both modes approach a vacuum-like state with the quantum fields exhibiting a stochastic broadening due to quantum noise.
\cref{fig:chaos_map opt 2,fig:chaos_map mech 2} coincide with the region displaying two stable fixed points in~\cref{fig:meanfield stability}.
Here, the optical mode is populated by a nearly-coherent state, as reported in~\cref{fig:delta_scan wigner 2}.
As such, the classical trajectory converges to one of the two lobes depending on the initial condition, while the quantum trajectories switch between the two metastable nearly-coherent states.
Conversely, the mechanical mode remains close to the vacuum at both the quantum and the classical level.
In this regime, the optomechanical setup effectively behaves similarly to the standard two-photon driven Kerr resonator~\cite{bartolo_exact_2016, minganti_spectral_2018, beaulieu_observation_2025}. 
A qualitatively different behavior is found across the classical region with zero stable fixed points.
\cref{fig:chaos_map opt 3,fig:chaos_map mech 3} show classical and quantum trajectory for a point with $\Lambda_{\rm max}=0$.
Both modes become macroscopically populated and the classical dynamics for $\alpha(t)$ and $\beta(t)$ flow in two limit cycles, as opposed to the quantum trajectories that display chaotic-like attractors instead.
This can be explained by the action of quantum jumps, which randomly force the quantum state out of the classical manifold~\cite{ferrari_dissipative_2025}.
Finally,~\cref{fig:chaos_map opt 4,fig:chaos_map mech 4} correspond to a point in~\cref{fig:chaos_map lyapunov} with $\Lambda_{\rm max}>0$ and $\alpha(t)$ and $\beta(t)$ flow into a strange attractor.
This behavior is mirrored by the quantum fields $\bra{\Psi(t)}\hat{a}\ket{\Psi(t)}$ and $\bra{\Psi(t)}\hat{b}\ket{\Psi(t)}$, where the stochastic action of the quantum jump amplifies chaotic oscillations.

\subsubsection{Spectral statistics of quantum trajectories}\label{sec:ssqt}

Finally, we investigate quantum signatures of dissipative chaos emerging from the Liouvillian spectrum.
To do so, we employ the spectral statistics of quantum trajectories (SSQT) criterion developed in Ref.~\cite{ferrari_dissipative_2025}.
The Liouvillian $\mathcal{L}$ can be diagonalized to obtain left and right eigenoperators
\begin{equation}
\mathcal{L}\hat{\eta}_j=\lambda_j\hat{\eta}_j,
\qquad\qquad
\mathcal{L}^\dagger\hat{\sigma}_j=\lambda_j^*\hat{\sigma}_j,
\end{equation}
that satisfy the bi-orthonormality relation $\operatorname{Tr}(\hat{\sigma}^\dagger_j\hat{\eta}_k)=\delta_{jk}$.
We then evolve single quantum trajectories $\hat{\rho}(t)=\ketbra{\Psi(t)}$ and decompose them over the eigenbasis of the Liouvillian
\begin{equation}\label{eq: liouvillian_decomposition}
    \hat{\rho}(t)
    =
    \sum_j c_j(t)\hat{\eta}_j,
    \qquad
    c_j(t)=\mathrm{Tr}\!\left[\hat{\sigma}_j^\dagger\rho(t)\right].
\end{equation}
The Liouvillian eigenvalues and eigenstates relevant for the trajectory's dynamics are those for which $|c_j(t)|>c_{\rm min}$ where $c_{\rm min}$ is a suitably chosen cutoff.
Additional details on the SSQT criterion can be found in Appendix~\ref{app:ssqt}.
Finally, we point out that given the size of the Liouvillian matrices we are considering, and to ensure convergence of the steady state and related observables, we can consider values of $G/\Omega_m$ corresponding to classical limit cycles but not chaotic attractors.
The latter occur indeed at parameters for which diagonalizing the entire Liouvillian becomes unfeasible. 

In~\cref{fig:spettri} we plot the bare Liouvillian eigenvalues in gray and highlight the set of those which are ``activated" according to the SSQT criterion over a single quantum trajectory.
In the presence of a single stable fixed point, only few eigenvalues close to the steady states are involved in the dynamics of $\ket{\Psi(t)}$ (cf.~\cref{fig:spettro 1}). This is consistent with the nearly-vacuum state encoded by $\hat{\rho}_{\rm ss}$ displayed in~\cref{fig:delta_scan wigner 1,fig:delta_scan wigner 2} for the optical mode and unraveled in~\cref{fig:chaos_map opt 1,fig:chaos_map mech 1} for both modes.
In the presence of two stable fixed points, a few more eigenvalues get selected by the SSQT criterion (see~\cref{fig:spettro 2}), and are always close to $\lambda_0=0$.
Analogously to the previous case, this is in line with the regular dynamics presented in~\cref{fig:chaos_map opt 2,fig:chaos_map mech 2}.
The picture is qualitatively different as we enter in the region with zero stable fixed points.
Here, thousands of eigenvalues contribute to the dynamics of a single quantum trajectory.
Moreover, they describe physical processes with an oscillation time larger than the decay rate, $\textrm{Im}[\lambda_j]/\textrm{Re}[\lambda_j]\gtrsim1$. 
Consequently, $\ket{\Psi(t)}$ is constituted by many rapidly oscillating processes, a picture consistent with an emergent complexity in the dynamics. 

Finally, we study the structure of the eigenstates selected by the SSQT analysis.
First, we quantify the support of $\ket{\Psi(t)}$ across the Liouvillian eigenbasis.
To this purpose, we expand $\hat{\rho}(t) = \ketbra{\Psi(t)}$ in the basis of the left eigenstates,
\begin{equation}
    \hat{\rho}(t)
    =
    \sum_j d_j(t)\hat{\sigma}^\dagger_j,
    \qquad d_j(t)=\mathrm{Tr}\!\left[\rho(t)\hat{\eta}_j\right].
\end{equation}
We then consider the quasi-probabilities~\cite{richter2025localizationdelocalizationquantumtrajectories}
\begin{equation}
    p_j = c_jd_j.
\end{equation}
Notably, while the SSQT implicitly relies on the normalization $\operatorname{Tr}(\hat{\eta}_j^\dagger\hat{\eta}_j)=1$, the quantities $p_j$ are independent of any normalization of the right eigenstates~\cite{richter2025localizationdelocalizationquantumtrajectories}.

In~\cref{fig:entropie 1} we plot the histogram of the quasi-probabilities $|p_j|$ for the three cases analyzed in~\cref{fig:spettri}.
We observe how the histograms corresponding to the classically regular regions with a finite number of stable fixed points are peaked below $10^{-20}$ (green curve, one stable fixed point) and $10^{-10}$ (orange curve, two stable fixed point), respectively.
This signals a low degree of wave function delocalization in the Liouvillian spectrum.
The histogram of $|p_j|$ corresponding to zero stable fixed points (blue curve) is instead peaked below $10^{-5}$, with many eigenstates associated with $|p_j|\sim 10^{-2}-10^{-1}$.
In this case, the quantum trajectory is delocalized across the spectrum, in line with~\cref{fig:spettro 3}.
These findings further confirm the predictions of the SSQT criterion.

To further characterize this spectral delocalization, we analyze the entropy of the individual eigenstates.
For each $\hat{\eta}_j$, we write the Hermitian and traceless combination $(\hat{\eta}_j+\hat{\eta}^\dagger_j)/2$ as $\hat{\rho}_j^+ - \hat{\rho}_j^-$.
Upon normalization, $\hat{\rho}_j^{\pm}$ are density matrix with the well defined Von Neumann entropy
\begin{equation}
    S_j^{\pm} = -\operatorname{Tr}\left[\hat{\rho}_j^{\pm}\log\hat{\rho}_j^{\pm}\right].
\end{equation}
Although the eigenstates of closed quantum systems are pure, their entanglement entropy provides a signature of quantum chaos and thermalization~\cite{dalessio_quantum_2016}.
In contrast, the states of open quantum systems are intrinsically mixed as the environment is traced out.
Therefore, a large entropy of many ``activated" chaotic modes should be linked to chaotic behavior in the system.
Results are plotted in~\cref{fig:entropie 2}.
The single- and two-stable fixed point examples exhibit a relatively low entropy on a small number of activated eigenstates. 
Conversely, the blue curve corresponding to classical limit cycle is characterized by highly entropic eigenstates selected by the SSQT, the number of which exceeds the classically regular case by more than two orders of magnitude.
This implies that $\ket{\Psi(t)}$ is delocalized over a massive amount of structurally complex eigenstates, suggesting that the classical instability is driven by dissipative chaotic quantum dynamics.

\section{Experimental parameters}\label{sec:exp_parameters}

\begin{figure*}[t!]
    \centering
    \includegraphics{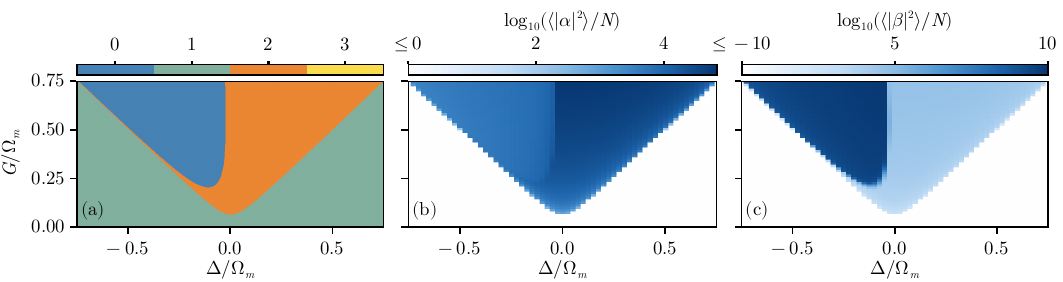}
    \caption{
    Mean-field stability diagram and population maps for the experimentally inspired parameters listed in~\cref{tab:real_param}, with nonlinear two-photon loss fixed to $\tilde{\kappa}_2/\Omega_m = 10^{-5}$.
    (a) Stability diagram in the $(\Delta/\Omega_m,G/\Omega_m)$ plane. The color code indicates the number of physical and stable stationary mean-field solutions. Compared with~\cref{fig:meanfield stability}, the same overall phase-boundary structure is retained, apart from the bistable region that is now absent on the right side.
    (b) Time-averaged optical population $\log_{10}(\langle|\alpha|^2\rangle/N)$, computed from the asymptotic time window after the transient dynamics has decayed. The optical field becomes populated inside the central triangular region delimited by the parametric threshold. Within this region, the population is larger in the region with two stable solutions (orange region in panel (a)), whereas the time-dependent region (blue region in panel (a)) displays a lower average population.
    (c) Time-averaged mechanical population $\log_{10}(\langle|\beta|^2\rangle/N)$, computed over the same time window. In contrast to the optical population, the phonon population is strongly enhanced in the time-dependent region, consistently with the mechanical response mechanism discussed in~\cref{sec:exp_parameters}. 
    }
    \label{fig:realparam}
    {\phantomsubcaptionlabel{fig:stability highK2}}
    {\phantomsubcaptionlabel{fig:phot highK2}}
    {\phantomsubcaptionlabel{fig:phon highK2}}
\end{figure*}

In all the previous sections, unless stated otherwise, the parameter values are set to the values reported in~\cref{fig:meanfield}.
In particular, the analysis performed so far employs an enhanced optomechanical coupling $g$, which keeps the photon and phonon populations within a range accessible to full quantum simulations.

We now focus instead on the semiclassical dynamics obtained using experimentally inspired parameters, in particular those reported in Ref.~\cite{Youssefi2023SqueezedMechanicalOscillator}.
The parameters used for this comparison are summarized in~\cref{tab:real_param}.
\begin{table}[h]
\centering
\renewcommand{\arraystretch}{1.25}
\begin{tabular*}{\columnwidth}{@{\extracolsep{\fill}} c c c}
\hline
Parameter & Value & Value in units of $\Omega_m$ \\
\hline
$\kappa_1$ & $\kappa_1/2\pi = 250~\mathrm{kHz}$ & $\kappa_1/\Omega_m = 0.139$ \\
$\tilde{g}$ & $\tilde{g}/2\pi = 13.4~\mathrm{Hz}$ & $\tilde{g}/\Omega_m = 7.44\times 10^{-6}$ \\
$\Gamma_m$ & $\Gamma_m/2\pi = 0.45~\mathrm{Hz}$ & $\Gamma_m/\Omega_m = 2.50\times 10^{-7}$ \\
\hline
\end{tabular*}
\caption{
Experimentally inspired parameters used as a benchmark for~\cref{fig:realparam}. The optical decay rate and optomechanical coupling are taken from Ref.~\cite{Youssefi2023SqueezedMechanicalOscillator}, with normalization $\Omega_m/2\pi=1.8~\mathrm{MHz}$. 
The mechanical damping is increased by one order of magnitude with respect to the experimental value to reduce the relaxation time and make the mean-field simulations tractable.
}
\label{tab:real_param}
\end{table}

Since the model considered here also includes a two-photon loss channel, we set its rate to $\tilde{\kappa}_2/\Omega_m = 10^{-5}$ for this comparison.
The resulting mean-field stability diagram and population maps are shown in \cref{fig:stability highK2,fig:phot highK2,fig:phon highK2}.

\Cref{fig:stability highK2} shows that the structure of the stability diagram is retained for this parameter set. 
The main difference is the disappearance of the bistable region, while the region with no stable mean-field solutions remains present. 
As discussed in \cref{sec:chaos}, this region is associated with limit cycles and chaotic strange attractors.

The optical population shown in \cref{fig:phot highK2} is largest in the region where two symmetry-related stable mean-field solutions exist, while the time-dependent region displays a smaller time-averaged optical population.
The mechanical population, shown in \cref{fig:phon highK2}, behaves differently: the phonon population is strongly enhanced precisely in the limit-cycles region, where the optical field undergoes persistent temporal oscillations.
This contrast shows that the mechanical occupation is not controlled only by the mean optical population, but also by the frequency content of the radiation-pressure force.
We make this point explicit by analyzing the mechanical response to the different Fourier components of the optical intensity.

To this end, we first consider \cref{eq:eoms_beta},
\begin{equation}
    \dot{\beta}+\left(i\Omega_m+\frac{\Gamma_m}{2}\right)\beta=ign_a(t),
    \label{eq:photresp 1}
\end{equation}
where $n_a(t)=|\alpha(t)|^2$ is the optical population. We decompose $n_a(t)$ into its time-averaged component and its finite-frequency components, $n_a(t)=\overline{n_a}+\sum_{\omega\neq0}n_\omega e^{-i\omega t}$, where $\overline{n_a}$ denotes the time average of the optical population over a sufficiently long time window. For a limit cycle, averaging over one oscillation period is sufficient.

Since Eq.~\eqref{eq:eoms_beta} is linear, the mechanical response can be decomposed into the same frequency components as the optical intensity,
\begin{equation}
    \beta(t)=\beta_0+\sum_{\omega\neq0}\beta_\omega e^{-i\omega t}.
    \label{eq:betaDecomposition}
\end{equation}
The static component $\beta_0$ is obtained by taking the time average of \cref{eq:photresp 1}. In the long-time periodic regime, $\overline{\dot{\beta}}=0$, $\overline{n_a(t)}=\overline{n_a}$, and $\overline{\beta(t)}=\beta_0$, which gives
\begin{equation}
    |\beta_0|^2=\frac{g^2\overline{n_a}^2}{\Omega_m^2+\Gamma_m^2/4}
\end{equation}
For the oscillating components we can consider a single Fourier component at frequency $\omega$,
$n_a(t)=n_\omega e^{-i\omega t}$, and the corresponding mechanical response,
$\beta(t)=\beta_\omega e^{-i\omega t}$. Taking the time derivative gives
$\dot{\beta}=-i\omega \beta_\omega e^{-i\omega t}$. Therefore, the mechanical equation becomes
\begin{equation}
\left[\frac{\Gamma_m}{2}+i(\Omega_m-\omega)\right]\beta_\omega
=
ign_\omega.
\end{equation}
Defining the mechanical susceptibility $\chi(\omega)=\left[\frac{\Gamma_m}{2}+i(\Omega_m-\omega)\right]^{-1}$ we get $\beta_\omega=ig\chi(\omega)n_\omega$. This yields
\begin{equation}
\begin{split}
&|\beta_\omega|^2=g^2|\chi(\omega)|^2|n_\omega|^2 \\ & |\chi(\omega)|^2=\frac{1}{(\Omega_m-\omega)^2+\Gamma_m^2/4}.
\end{split}
\end{equation}
Therefore, each finite-frequency component of $n_a(t)$ contributes to the mechanical response with a weight set by $|\chi(\omega)|^2$.
The phonon number is $n_b(t)=|\beta(t)|^2$. Using~\cref{eq:betaDecomposition}, we obtain
\begin{equation} 
\begin{split}
|\beta(t)|^2=|\beta_0|^2&+\sum_{\omega\neq0}\left[\beta_0\beta_\omega^*e^{i\omega t}+\beta_0^*\beta_\omega e^{-i\omega t}\right]\\&+\sum_{\omega,\omega'\neq0}\beta_\omega\beta_{\omega'}^*e^{i(\omega-\omega')t}.
\end{split}
\end{equation}
Upon time averaging, the oscillating terms vanish, while the diagonal terms satisfy
$\overline{e^{i(\omega-\omega')t}}=\delta_{\omega,\omega'}$. Therefore,
\begin{equation}
    \begin{aligned}
    \overline{n_b}&=\overline{|\beta(t)|^2}
    =|\beta_0|^2+\sum_{\omega\neq0}|\beta_\omega|^2\\
    &=\frac{g^2}{\Omega_m^2+\frac{\Gamma_m^2}{4}}\overline{n_a}^2+\sum_{\omega\neq0}\frac{g^2}{(\Omega_m-\omega)^2+\frac{\Gamma_m^2}{4}}|n_\omega|^2 .
    \end{aligned}
\end{equation}
The last equation shows that the average phonon number depends on both the static optical population $\overline{n_a}$ and the finite-frequency optical components $|n_\omega|^2$. The latter are weighted by the mechanical susceptibility $|\chi(\omega)|^2$. In the low-dissipation limit, $\Gamma_m\ll\Omega_m$, this susceptibility is sharply peaked around the mechanical resonance. Therefore, the dominant contribution to the sum comes from optical-intensity components with frequencies close to $\Omega_m$. Keeping only the resonant contribution, the average phonon number reduces to
\begin{equation}
    \overline{n_b}\simeq\frac{g^2\overline{n_a}^2}{\Omega_m^2+\Gamma_m^2/4}+\frac{4g^2}{\Gamma_m^2}|n_{\Omega_m}|^2 .
    \label{eq:phononNumber}
\end{equation}

This expression explains the behavior of the phonon population shown in \cref{fig:phon highK2}. In the limit-cycles region, the optical intensity is not stationary and contains finite-frequency components. When components close to the mechanical resonance are present, they are resonantly amplified by the mechanical susceptibility, leading to a large increase of the time-averaged phonon population. By contrast, in the stable fixed-point regions the optical population is stationary, so that the finite-frequency contribution is absent and the mechanical population is determined only by the static term.

For the experimentally inspired parameters considered here, the hierarchy $\Omega_m\gg g\gg\Gamma_m$ makes this distinction particularly pronounced. The static contribution is suppressed by the large mechanical frequency, scaling approximately as $g^2/\Omega_m^2$, whereas the resonant contribution scales as $4g^2/\Gamma_m^2$. Therefore, whenever the optical dynamics contains a component close to $\Omega_m$, the oscillating contribution can dominate the mechanical population.

Overall, this comparison shows that the semiclassical regimes analyzed throughout the paper are not tied to the enhanced optomechanical coupling used in the full quantum simulations. When experimentally inspired optomechanical parameters are used, the same mean-field structure is recovered, although at photon and phonon populations that are several orders of magnitude larger.

\section{Conclusion}\label{sec:conclusion}

We have studied a two-photon-driven quantum optomechanical system in which a nonlinear optical resonator is coupled by radiation pressure to a mechanical oscillator and evolves in the presence of single-photon, two-photon, and mechanical losses. The two-photon drive plays a central role: it preserves the discrete transformation $\hat{a}\to-\hat{a}$ and therefore endows the Liouvillian dynamics with a weak $\mathbb{Z}_2$ symmetry. This symmetry, together with the nonlinear optomechanical feedback, allows critical and chaotic driven-dissipative phenomena to emerge in the same microscopic model.

Our first main result concerns dissipative phase transitions with spontaneous symmetry breaking. At the semiclassical level, the system displays vacuum-like, symmetry-broken, bistable, and dynamically unstable regions. The low-lying Liouvillian spectrum confirms that the negative-detuning transition is second order: the steady-state populations grow continuously, and an antisymmetric mode closes throughout the symmetry-broken phase in the thermodynamic limit. At positive detuning the transition becomes first order, with a symmetric mode approaching zero near the critical point and the coexistence of low- and high-occupation structures.

Our second main result concerns the emergence of classical chaos with consistent signatures of quantum chaotic dissipative dynamics~\cite{sa2026talktalkdissipativequantum}. For stronger two-photon pumping, the mean-field analysis reveals a broad region without stable fixed points. 
There, the classical dynamics develops regular limit cycles and, deeper in parameter space, strange attractors with positive largest Lyapunov exponent. 
The corresponding quantum trajectories inherit this time-dependent structure while being broadened by quantum jumps and fluctuations, and delocalize over many strongly entropic Liouvillian eigenoperators. 

Taken together, these results provide a proof of principle that driven-dissipative optomechanical setups can host emergent physics beyond static bistability, including spontaneous symmetry breaking, dissipative criticality, and chaotic dynamics. This is particularly relevant because optomechanical systems naturally connect optical fields to massive mechanical degrees of freedom. Two-photon driving can therefore be used not only to engineer protected optical parity sectors, but also to imprint symmetry-broken and dynamically complex states onto quantum mechanical motion.

This perspective is important for future tabletop experiments aimed at probing the foundations of quantum mechanics with increasingly macroscopic mechanical systems. The symmetry-broken states identified here are separated in phase space and coupled to mechanical displacement, making them a natural starting point for protocols that generate, stabilize, and detect superpositions of macroscopically distinct mechanical configurations.

\begin{acknowledgments}
We acknowledge enightening discussions with Matteo Fadel.
A.M., F.F., L.F., M.S.\@ and V.S.\@ acknowledge support from the Swiss National Science Foundation through Projects No. 200020\_215172, 200021-227992, and 20QU-1\_215928, and as a part of NCCR SPIN (grant number 225153).
A.M. acknowledges the Swiss Quantum Initiative (SQI) of the Swiss Academy of Sciences (SCNAT) through the 2024 Quantum Voucher Model for the Grant/Ruling 24\_1084.
M.S.\@ acknowledges funding from the Swiss Academy of Sciences (SCNAT) through the Swiss Quantum Initiative (SQI) Grant No.\@ 24\_1111.
V.M.\@ acknowledges PNRR MUR project, National Quantum Science and Technology Institute (NQSTI) Grant No. PE0000023. 
\end{acknowledgments}

\appendix

\begin{widetext}

\section{\label{app: physicality and stability}Physicality and Stability}

In \cref{sec: semiclassApp} we presented the stationary solutions of the mean-field equations and their stability regimes. Starting from \cref{eq:eoms_alpha,eq:eoms_beta} a complete derivation can be performed. The steady state solutions of the system are found by imposing the stationary conditions $\dot\alpha =0$ and $\dot\beta=0$.
Solving, we find the condition on $\beta$
\begin{equation}
\begin{split}
    \beta_s =
    \frac{i g |\alpha|^2}{i\Omega_m+\Gamma_m/2}
    \qquad\Longrightarrow\qquad
    \beta_s+\beta_s^*=2 \mathrm{Re}[\beta]=\frac{U_{\rm eff}}{g}\lvert\alpha\rvert^2
\end{split}
\end{equation}
where $U_{\rm eff}$ is given by
\begin{equation}
    U_{\rm eff}=\frac{2g^2\Omega_m}{\Omega_m^2+\left(\frac{\Gamma_m}{2}\right)^2}
    \label{eq: real beta2}
\end{equation}
Using this condition on the mechanical field and $\dot\alpha=0$ the first mean-field equation reduces to
\begin{equation}
     \left(-i\Delta - \frac{\kappa_1}{2}\right)\alpha + i\,U_{\rm eff}\,\lvert\alpha\rvert^2\alpha - \kappa_2 |\alpha|^2 \alpha - i G \alpha^*=0,
     \label{eq: stationary alpha2}
\end{equation}
which coincides with the equation of a two-photon driven Kerr resonator with an effective nonlinearity $U_{\rm eff}$.
To solve this equation $\alpha$ can be written in complex form as $\alpha=\sqrt{n} e^{i\phi}$, where $n$ is the number of photons in the cavity:
\begin{equation}
    \left[-i(\Delta-U_{\rm eff}n)-\frac{\kappa_1}{2}-\kappa_2n\right]\sqrt{n}e^{i \phi}=iG\sqrt{n}e^{-i \phi}
    \label{eq: complete}
\end{equation}
This equation always admits $n=0$ as a solution. 
This solution is always physical and represents the situation with an empty cavity and mechanical mode. 
To find the other solutions we take the modulus of \cref{eq: complete}:
\begin{equation}
    n^2(U_{\rm eff}^2+\kappa_2^2)+n(\kappa_1\kappa_2-2\Delta U_{\rm eff})+\Delta^2+\left(\frac{\kappa_1}{2}\right)^2=F^2 
\label{eq: modulus}
\end{equation}
Solving this $\phi-$independent equation for $n$ yields two possible results:
\begin{equation}
    n^{(\pm)}_a =
    \frac{
    -\kappa_1\kappa_2/2+U_{\rm eff}\Delta}{U_{\rm eff}^2+\kappa_2^2}
    \pm
    \frac{\sqrt{
    (U_{\rm eff}^2+\kappa_2^2)G^2
    -
    \left(\kappa_2\Delta+U_{\rm eff}\kappa_1/2\right)^2
    }}
    {U_{\rm eff}^2+\kappa_2^2}.
\end{equation}
Going back to \cref{eq: complete} and setting $n=n_s$, where $n_s$ is one of the two possible finite-amplitude solutions, gives the following equation for the phase $\phi_s$ of the stationary solution:
\begin{equation}
    e^{2i\phi_s}=-\frac{(\Delta-U_{\rm eff}n_s)+i\left(\kappa_1/2+\kappa_2n_s\right)}{G}
    \label{eq: phase}
\end{equation}
From \cref{eq: phase} it can be seen that, once $n=n_s$ is set, both $\phi=\phi_s$ and $\phi=\phi_s + \pi$ satisfy the relation.
Therefore, five stationary solutions are possible: the vacuum $n=0$, and four paired solutions $(n_+,\phi_+), (n_+,\phi_++\pi), (n_-,\phi_-),  (n_-,\phi_-+\pi)$.
These solutions are mathematically possible; however, they are physically meaningful (real and positive $n$) only within specific regions of the parameter space. 
To determine when the last four solutions are physical we consider again \cref{eq: modulus}.
The left-hand side represents an upward-opening parabola in the $n$--$G$ plane ($G>0$), with vertex $(n^*,G_{\mathrm{min}})$ and intercept on the $G$-axis $(0,G_\mathrm{thr})$:
\begin{equation}
n^{(*)} = \frac{2U_{\rm eff}\Delta - \kappa_1\kappa_2}{2(U_{\rm eff}^2 + \kappa_2^2)},\qquad G_\mathrm{min} = \sqrt{\frac{\left(\Delta\kappa_2 + U_{\rm eff}\frac{\kappa_1}{2}\right)^2}{U_{\rm eff}^2 + \kappa_2^2}},\qquad G_\mathrm{thr} = \sqrt{\Delta^2 + \left(\frac{\kappa_1}{2}\right)^2}
\end{equation}
Using these points, together with geometrical considerations, it is possible to identify the regions in the ($\Delta/\Omega_m, \,G/\Omega_m$) plane where the solutions are physical. 
It is now important to determine whether such solutions are stable. 
This can be done systematically by deriving the drift matrix $M_{\alpha\alpha^*\beta\beta^*}$ of the system. The general drift matrix on the $(\alpha, \alpha^*, \beta , \beta^*)$ four-dimensional plane contains complex entries, therefore it is convenient to move to the quadrature basis.
For the cavity field the new basis is $x_c=\sqrt{2}\textrm{Re}(\alpha)$ and $p_c=\sqrt{2}\textrm{Im}(\alpha)$, while for the mechanical resonator $x_m=\sqrt{2}\textrm{Re}(\beta)$ and $p_m=\sqrt{2}\textrm{Im}(\beta)$. In this representation, the drift matrix becomes real and takes the form
    \begin{equation}
\resizebox{\textwidth}{!}{$
M_j = \begin{pmatrix}
-\dfrac{\kappa_1}{2} - 2\kappa_2|\mathrm{Re}(\alpha_j)|^2 - \kappa_2|\alpha_j|^2 &
\Delta - 2g\,\mathrm{Re}(\beta_j) - G - 2\kappa_2\,\mathrm{Re}(\alpha_j)\,\mathrm{Im}(\alpha_j) &
-2g\,\mathrm{Im}(\alpha_j) & 0 \\[8pt]

-\Delta + 2g\,\mathrm{Re}(\beta_j) - G - 2\kappa_2\,\mathrm{Re}(\alpha_j)\,\mathrm{Im}(\alpha_j) &
-\dfrac{\kappa_1}{2} - \kappa_2|\alpha_j|^2 - 2\kappa_2|\mathrm{Im}(\alpha_j)|^2 &
2g\,\mathrm{Re}(\alpha_j) & 0 \\[8pt]

0 & 0 & -\dfrac{\Gamma_m}{2} & \Omega_m \\[8pt]

2g\,\mathrm{Re}(\alpha_j) & 2g\,\mathrm{Im}(\alpha_j) & -\Omega_m & -\dfrac{\Gamma_m}{2}
\end{pmatrix}
$}
\label{eq: driftmatrix}
\end{equation}
In particular, for each steady state solution, labeled by $j=1,2,3,4,5$, the stability can be analyzed by substituting its steady-state values:
\begin{equation}
\left\{
\begin{array}{l}
\mathrm{Re}(\alpha_j) = \sqrt{n_j}\cos(\phi_j) \\
\mathrm{Im}(\alpha_j) = \sqrt{n_j}\sin(\phi_j) \\
\mathrm{Re}(\beta_j) = 
\dfrac{C}{2g}\,|\alpha_j|^2
\end{array}
\right.
\end{equation}
A solution is stable if and only if all the eigenvalues of the associated drift matrix have a negative real part. This condition is equivalently satisfied when the Routh-Hurwitz criterion holds \cite{DeJesus1987RouthHurwitz}.
In order to apply the Routh-Hurwitz criterion, the coefficients of the characteristic polynomial of the drift matrix $M_j$ and the matrix itself are required. Given the characteristic polynomial $p(\lambda)=a_4\lambda^4+a_3\lambda^3+a_2\lambda^2+a_1\lambda+a_0$ the coefficients are:

\begin{equation}
\begin{split}
a_0 &= 
\left(\frac{1}{16}\Gamma_m^2 + \frac{1}{4}\Omega_m^2\right)
\Bigl(
-4G^2 + \kappa_1^2 + 12\kappa_1\kappa_2 n_j
+ 4\bigl((C^2 + 5\kappa_2^2)n_j^2 - 2 C n_j \Delta + \Delta^2
\Bigr)
+ 8 g^2 n_j (C n_j - \Delta)\Omega_m, \\[6pt]
a_1 &= 
- G^2 \Gamma_m 
+ \frac{1}{4}\Gamma_m\bigl(\kappa_1^2 + \kappa_1(12\kappa_2 n_j + \Gamma_m)\bigr)
+ 4\bigl(C^2 n_j^2 + \kappa_2 n_j (5\kappa_2 n_j + \Gamma_m)
- 2 C n_j \Delta + \Delta^2\bigr)
+ (\kappa_1 + 4\kappa_2 n_j)\Omega_m^2, \\[6pt]
a_2 &= 
- G^2 + \frac{\kappa_1^2}{4} 
+ C^2 n_j^2 + 5\kappa_2^2 n_j^2 
+ 4\kappa_2 n_j \Gamma_m
+ \frac{\Gamma_m^2}{4}
+ \kappa_1(3\kappa_2 n_j + \Gamma_m)
- 2C n_j \Delta + \Delta^2 + \Omega_m^2, \\[6pt]
a_3 &= \kappa_1 + 4\kappa_2 n_j + \Gamma_m.
\end{split}
\end{equation}
\end{widetext}
The coefficients obtained are phase-independent.
For a $4\times4$ matrix, a stationary solution is stable (all $M_j$ eigenvalues have a negative real part) if all the following conditions are satisfied (Routh-Hurwitz criterion):
\begin{equation}
\begin{split}
    &a_3 > 0, \quad a_2 > 0, \quad a_1 > 0, \quad a_0 > 0,\\
    &a_2 a_3 - a_1 > 0,\\
    &a_1 a_2 a_3 - a_1^2 - a_3^2 a_0 > 0
    \label{eq:relevant condition}
\end{split}
\end{equation}
For the solutions corresponding to $n = n_\pm$, the stability analysis can be performed numerically.  
This procedure yields the stability diagram presented in~\cref{fig:meanfield stability}, showing that the only relevant condition among the six listed in~\cref{eq:relevant condition} is the last one, as the others are always satisfied within the physical region.

For the $n = 0$ solution, however, the analysis can be done analytically and the coefficients are given by:
\begin{equation}
\begin{split}
    a_0 &= 
    \left(\frac{1}{4}\Gamma_m^2 + \Omega_m^2\right)
    \Bigl(
    -G^2 + \frac{\kappa_1^2}{4} +\Delta^2\Bigr) \\
    a_1 &= 
    \Gamma_m\Bigl(-G^2 + \frac{\kappa_1^2}{4} +\Delta^2+\frac{\Gamma_m}{4}\kappa_1\Bigr)+\Omega_m^2\kappa_1 \\
    a_2 &= 
    - G^2 + \frac{\kappa_1^2}{4} 
    + \frac{\Gamma_m^2}{4}
    + \kappa_1\Gamma_m
    + \Delta^2 + \Omega_m^2, \\
    a_3 &= \kappa_1 + \Gamma_m
\end{split}
\end{equation}
In this case, the first condition requires $G < G_\mathrm{thr}$, which is also sufficient to ensure that all the remaining Routh–Hurwitz inequalities are fulfilled, therefore $n=0$ is stable when the pumping power $G$ is below the parametric threshold $G_\mathrm{thr}$. 

\section{\label{app:more_on_classical_chaos}Additional details on the classical chaotic dynamics}
\begin{figure}[t!]
    \centering
    \includegraphics[width=\columnwidth]{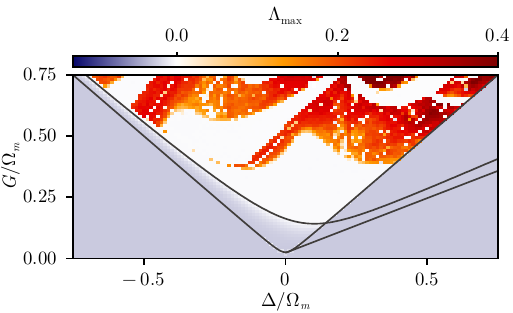}
    \caption{
    Lyapunov map obtained with the same procedure and parameters as in \cref{fig:chaos_map lyapunov}, but using the initial-condition amplitude sampled from $\mathcal{N}(0,0.1)$ instead of $\mathcal{N}(0,2)$.
    The side peak visible in \cref{fig:chaos_map lyapunov} is not reached, and the right-side coexistence region is dominated by relaxation to the stable empty solution.
    This comparison shows that, when different attractors coexist, $\Lambda_{\mathrm{max}}$ refers to the attractor selected by the chosen initial condition.
    } 
    \label{fig:caosLow}
\end{figure}

The Lyapunov map depends on the basin of attraction selected by the initial condition. 
This is shown in \cref{fig:caosLow}, where the same parameters as in \cref{fig:chaos_map lyapunov} are used, but the initial conditions are sampled from the complex Gaussian distribution $\mathcal{N}(0, 0.01)$. 
With this smaller sampling amplitude, the region hosting limit cycles and chaotic attractors shrinks to the zero-stable fixed point region predicted by the stability map in~\cref{fig:meanfield stability}.
Moreover, the dip around $\Delta/\Omega:m=-0.5$ visible in \cref{fig:chaos_map lyapunov} disappears.

\section{\label{app:quantum_trajectories}Stochastic quantum trajectories}

We review here the theory of quantum trajectories, an open system tool we broadly used in the main text. 
While the density matrix describes the ensemble-averaged evolution of the open system, the Monte Carlo wave function construction gives access to individual stochastic realizations of the same Lindblad dynamics \cite{BreuerPetruccione2002, dalibard_wave-function_1992}. Each realization is a pure-state trajectory
\begin{equation}
    \rho_m(t)
    =
    |\psi_m(t)\rangle\langle\psi_m(t)| ,
\end{equation}
and the density matrix is recovered by averaging over trajectories,
\begin{equation}
    \rho(t)
    =
    \overline{
    |\psi_m(t)\rangle\langle\psi_m(t)|
    },
\end{equation}
where \(m\) labels the trajectory and the overline denotes the ensemble average.
For a photon-counting unraveling, the state evolves between quantum jumps under the effective non-Hermitian Hamiltonian
\begin{equation}
    \hat{H}_{\mathrm{eff}}
    =
    \hat{H}
    -
    \frac{i}{2}
    \sum_{\mu}
    \hat{L}_{\mu}^{\dagger}\hat{L}_{\mu},
    \label{eq: Heff}
\end{equation}
where \(\hat{L}_{\mu}\) are the Lindblad jump operators. 
For the model studied in this article, they are
\begin{equation}
    \hat{L}_1=\sqrt{\kappa_1}\hat{a},
    \qquad
    \hat{L}_2=\sqrt{\kappa_2}\hat{a}^2,
    \qquad
    \hat{L}_3=\sqrt{\Gamma_m}\hat{b}.
    \label{eq: collapse_operators}
\end{equation}
During the non-unitary evolution generated by \(\hat{H}_{\mathrm{eff}}\), the norm of the wave function decreases. For a small time interval \(dt\), this norm loss determines the total probability for a jump to occur,
\begin{equation}
    dp
    =
    dt
    \sum_{\mu}
    \langle \psi_m(t)|
    \hat{L}_{\mu}^{\dagger}\hat{L}_{\mu}
    |\psi_m(t)\rangle .
    \label{eq: jump_probability}
\end{equation}
If no jump occurs, the state is propagated with \(\hat{H}_{\mathrm{eff}}\) and then normalized. If instead a jump occurs in the channel \(\mu\), the state is updated as
\begin{equation}
    |\psi_m(t)\rangle
    \longrightarrow
    \frac{
    \hat{L}_{\mu}|\psi_m(t)\rangle
    }{
    \sqrt{
    \langle \psi_m(t)|
    \hat{L}_{\mu}^{\dagger}\hat{L}_{\mu}
    |\psi_m(t)\rangle
    }
    } .
    \label{eq: quantum_jump}
\end{equation}
When several collapse channels are present, the probability that the jump is associated with channel \(\mu\) is
\begin{equation}
    P_{\mu}(t)
    =
    \frac{
    \langle \psi_m(t)|
    \hat{L}_{\mu}^{\dagger}\hat{L}_{\mu}
    |\psi_m(t)\rangle
    }{
    \sum_{\nu}
    \langle \psi_m(t)|
    \hat{L}_{\nu}^{\dagger}\hat{L}_{\nu}
    |\psi_m(t)\rangle
    }.
    \label{eq: jump_channel_probability}
\end{equation}
We used the standard Monte Carlo unraveling in~\cref{sec:phasespace}.

In~\cref{sec:chaos}, we instead considered a different unraveling: the \(\beta\)-dyne unraveling \cite{Minganti2022BetaDyne}. 
This displacement is applied to the monitored single-photon optical loss channel. Defining
\begin{equation}
    \hat{L}_1=\sqrt{\kappa_1}\hat{a},
\end{equation}
the collapse operator is replaced by
\begin{equation}
    \hat{L}_1
    \longrightarrow
    \hat{L}_{1,\beta}
    =
    \hat{L}_1+\beta ,
    \label{eq: beta_dyne_L}
\end{equation}
and the Hamiltonian is shifted as
\begin{equation}
    \hat{H}
    \longrightarrow
    \hat{H}_{\beta}
    =
    \hat{H}
    -
    \frac{i}{2}
    \beta
    \left(
    \hat{L}_1-\hat{L}_1^{\dagger}
    \right).
    \label{eq: beta_dyne_H}
\end{equation}The collapse operators are re-defined as
\begin{equation}
    \hat{L}_{1,\beta}
    =
    \sqrt{\kappa_1}\hat{a}+\beta,
    \quad
    \hat{L}_2=\sqrt{\kappa_2}\hat{a}^2,
    \quad
    \hat{L}_3=\sqrt{\Gamma_m}\hat{b}.
    \label{eq: beta_collapse_operators}
\end{equation}
In all the simulations involving this kind of unraveling, we considered $\beta=2$. 

In the presence of a weak $\mathbb{Z}_2$ symmetry, such an unraveling is necessary as otherwise $\bra{\Psi(t)}\hat{a}\ket{\Psi(t)}=0$ for $\beta=0$ (i.e., photon-counting unraveling).
A nonzero $\beta$ allows visualizing the quantum dynamics in phase space for the complex quadratures.
More details about this point can be found in Ref.~\cite{Minganti_2018}.

\section{Additional details on the SSQT criterion}\label{app:ssqt}
We provide here addition al information about the SSQT criterion employed in~\cref{sec:chaos}.
To determine the Liouvillian eigenvalues and eigenstates involved in the dynamics of single quantum trajectories, we consider the stochastic realization $m$ at fixed time $t$ and the spectral coefficient $c_{m,j}(t)$ in~\cref{eq: liouvillian_decomposition} (notice that here we introduced the index $m$ counting the quantum trajectories).
To distinguish eigenstates contributing to the structure of $\ket{\Psi_m(t)}$ from those yielding a negligible contribution, we introduce a characteristic scale of the spectral weights. 
This is defined as the center of mass of the coefficient distribution~\cite{ferrari_dissipative_2025}
\begin{equation}
    C_m(t)
    =
    \frac{
    \sum_j |c_{m,j}(t)|^2
    }{
    \sum_j |c_{m,j}(t)|
    }.
    \label{eq:ssqt_center_mass}
\end{equation}
If more than one trajectory is used, the corresponding scale is averaged over trajectories,
\begin{equation}
    \overline{C}(t)
    =
    \frac{1}{N_{\mathrm{traj}}}
    \sum_{m=1}^{N_{\rm traj}} C_m(t).
\end{equation}
The cutoff is then fixed as
\begin{equation}
    c_{\mathrm{min}}(t)
    =
    \overline{C}(t)\times10^{-k},
    \qquad
    k\in[2,4].
    \label{eq:ssqt_cutoff}
\end{equation}
Then, only eigenvalues and eigenstates such that $|c_{m,j}(t)|>c_{\mathrm{min}}(t)$ are considered.
The cutoff therefore selects the Liouvillian modes whose spectral weight is large compared with the typical scale of the coefficient distribution. 

\begin{figure}[t!]
    \centering
    \includegraphics[width=\columnwidth]{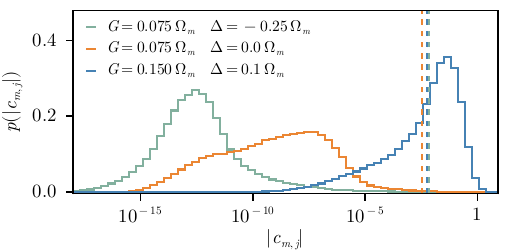}
    \caption{
    Distributions of the spectral coefficients $|c_{m,j}|$ used to select the active Liouvillian modes. For each parameter point, the histogram is obtained by collecting all coefficients from the stochastic trajectories and all Liouvillian modes, forming the set $\{|c_{m,j}|\,\,\forall m,j\}$, and then binning this full set. In the present figure, 10 stochastic trajectories are used for each parameter point. The vertical dashed lines indicate the corresponding cutoffs $c_{\mathrm{min}}$ defined in \cref{eq:ssqt_cutoff}. In the regular low-pump regimes, most coefficients lie below the cutoff and only a small number of modes is selected. In the semiclassical limit-cycle regime, the distribution shifts toward larger values of $|c_{m,\,j}|$, so that many more modes become active.
    }
    \label{fig:coeffChoice}
\end{figure}

\cref{fig:coeffChoice} shows the distributions $p(|c_{m,\,j}|)$ for the three representative parameter points analyzed in the main text. The vertical dashed lines indicate the corresponding values of $c_{\mathrm{min}}$. In the low-pump regular regimes, the distributions are concentrated mainly below the cutoff, so that only a small number of modes is selected. In contrast, in the semiclassical limit-cycle regime the distribution shifts toward larger values of $|c_{m,\,j}|$, and many more coefficients lie above the cutoff. This is consistent with the broad activation of Liouvillian modes shown in \cref{fig:spettri}.

The value of $k$ therefore fixes the numerical cutoff used to separate relevant and negligible spectral components. In the present analysis $k=2$ is set, corresponding to $c_{\mathrm{min}}=\overline{C}\times10^{-2}$. This choice is heuristic, but it provides a practical criterion to retain the Liouvillian modes with an appreciable projection on the stochastic trajectory while discarding modes with only negligible weight.

\section{Complex spacing ratios}\label{app: complexspacing}

\begin{figure}[t!]
    \centering
    \includegraphics[width=\columnwidth]{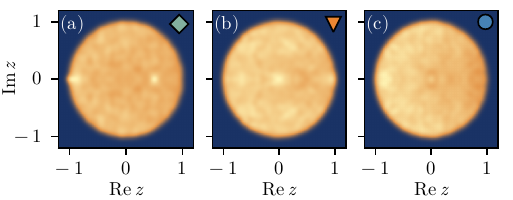}
\caption{
    Density of the complex spacing ratios $z_j$ in the complex plane for the three representative parameter points analyzed in the main text, with $N=1$.
    The panels correspond to: (a) $\Delta/\Omega_m=-0.25$, $G/\Omega_m=0.075$; (b) $\Delta/\Omega_m=0$, $G/\Omega_m=0.075$; (c) $\Delta/\Omega_m=0.1$, $G/\Omega_m=0.15$.
    }
    \label{fig:complexspacings}
\end{figure}

\begin{table}[t!]
\centering
\setlength{\tabcolsep}{10pt}
\renewcommand{\arraystretch}{1.25}
\begin{tabular}{c c c c c}
\hline
 & $\Delta/\Omega_m$ & $G/\Omega_m$ & $-\langle\cos\theta\rangle$ & $\langle r\rangle$ \\
\hline
(a) & $-0.25$ & $0.075$ & $0.0372$ & $0.6757$ \\
(b) & $0$ & $0.075$ & $0.0310$ & $0.6695$ \\
(c) & $0.1$ & $0.15$ & $0.0737$ & $0.6731$ \\
\hline
Poisson & -- & -- & $0$ & $2/3$ \\
Ginibre & -- & -- & $\simeq 0.24$ & $\simeq 0.738$ \\
\hline
\end{tabular}
\caption{
Complex-spacing-ratio indicators for the three representative parameter points. The quantity $-\langle\cos\theta\rangle$ measures the angular anisotropy of the distribution, while $\langle r\rangle$ measures the radial component.
}
\label{tab:complex_spacing}
\end{table}

In this Appendix we study the spectral statistics of the Liouvillian eigenvalues, a commonly used signature of dissipative quantum chaos.
Specifically, we analyze the complex spacing ratios introduced in Ref.~\cite{sa_complex_2020}. 
Given a complex Liouvillian eigenvalue $\lambda_j$, we identify its nearest neighbor $\lambda_j^{\mathrm{NN}}$ and next-nearest neighbor $\lambda_j^{\mathrm{NNN}}$ in the complex plane. 
The complex spacing ratio is then defined as
\begin{equation}
    z_j
    =
    \frac{
    \lambda_j^{\mathrm{NN}}-\lambda_j
    }{
    \lambda_j^{\mathrm{NNN}}-\lambda_j
    }
    =
    r_j e^{i\theta_j},
    \label{eq: complex_spacing_ratio}
\end{equation}
This quantity has the advantage of avoiding the unfolding procedure required by spacing-based spectral diagnostics \cite{akemann_universal_2019}. For uncorrelated complex eigenvalues, corresponding to the integrable case, the distribution of $z_j$ is expected to be approximately flat inside the unit circle. In contrast, for chaotic non-Hermitian spectra in the Ginibre universality class~\cite{ginibre_statistical_1965}, the distribution develops level repulsion around the origin and an angular anisotropy.

The two single-number indicators used to summarize the distribution are
\begin{equation}
    \langle r\rangle
    =
    \frac{1}{N_\lambda}
    \sum_j r_j,
    \qquad
    \langle \cos\theta\rangle
    =
    \frac{1}{N_\lambda}
    \sum_j \cos\theta_j .
\end{equation}
For uncorrelated complex eigenvalues one expects
\begin{equation}
    \langle r\rangle = \frac{2}{3},
    \qquad
    -\langle\cos\theta\rangle = 0,
\end{equation}
while for large Ginibre random matrices the expected values are approximately
\begin{equation}
    \langle r\rangle \simeq 0.738,
    \qquad
    -\langle\cos\theta\rangle \simeq 0.24 .
\end{equation}
Thus, an increase of $-\langle\cos\theta\rangle$ provides a signature of angular anisotropy, while an increase of $\langle r\rangle$ indicates stronger spectral correlations.

The results for the three parameter points analyzed in the main text are reported in~\cref{tab:complex_spacing}. The third point, corresponding to the semiclassical limit-cycle regime, is the one where signatures of dissipative quantum chaos are expected to be more visible. The value of $-\langle\cos\theta\rangle$ is indeed larger than in the two low-pump regimes, but the increase is weak. Similarly, $\langle r\rangle$ remains close to the Poisson value $2/3$ for all three points.

As discussed in \cref{sec:signatures}, the weak separation between the three cases could be attributed to the different energy scales of the problem, where $\Omega_m\gg g,\Gamma_m>\kappa_2$, with the dominant contribution coming from the weakly dissipative harmonic oscillator $\Omega_m\hat{b}^\dagger\hat{b}$.
We cannot exclude \textit{a priori} that larger drive amplitudes, deeper in the classical region with zero stable fixed points, would eventually lead to the emergence of Ginibre statistics in the spectrum.
Or, otherwise, that an alternative mechanism could explain the lack of spectral correlations.
Investigating these points would require a major computational effort, due to the growing cutoffs in the bosonic Hilbert space, that goes beyond the scope of this paper.

\bibliography{references}

\end{document}